\documentclass[useAMS,usenatbib]{mn2e}

\usepackage{graphicx}

\begin{document}


\newcommand{\lya}{Ly$\alpha$}
\newcommand{\caii}{\hbox{Ca\,{\sc ii}}}
\newcommand{\caiii}{\hbox{Ca\,{\sc iii}}}
\newcommand{\mgii}{\hbox{Mg\,{\sc ii}}}
\newcommand{\feii}{\hbox{Fe\,{\sc ii}}}
\newcommand{\znii}{\hbox{Zn\,{\sc ii}}}
\newcommand{\crii}{\hbox{Cr\,{\sc ii}}} 
\newcommand{\hi}{\hbox{H\,{\sc i}}} 
\newcommand{\hii}{\hbox{H\,{\sc ii}}}
\newcommand{\nhi}{$N$(\hbox{H\,{\sc i}})}
\newcommand{\nzn}{$N$(\hbox{Zn\,{\sc ii}})}
\newcommand{\ncr}{$N$(\hbox{Cr\,{\sc ii}})}
\newcommand{\ncaii}{$N$(\hbox{Ca\,{\sc ii}})}
\newcommand{\acm}{atoms\,cm$^{-2}$}
\newcommand{\wmg}{W$_0^{\lambda 2796}$}
\newcommand{\wca}{W$_0^{\lambda 3935}$}
\newcommand{\apjs}{ApJS}
\newcommand{\apj}{ApJ}
\newcommand{\aj}{AJ}
\newcommand{\araa}{ARA\&A}
\newcommand{\apjl}{ApJL}
\newcommand{\nat}{Nature}
\newcommand{\pasp}{PASP}
\newcommand{\aap}{A\&A}
\newcommand{\mnras}{MNRAS}

\def\ltsima{$\; \buildrel < \over \sim \;$}
\def\simlt{\lower.5ex\hbox{\ltsima}}
\def\gtsima{$\; \buildrel > \over \sim \;$}
\def\simgt{\lower.5ex\hbox{\gtsima}}


\title[Ca\,II, metals, and dust in DLAs]
{Measurements of Ca\,{\sc ii} absorption, metals, and dust in a sample of
  $z\simeq 1$ DLAs and subDLAs\thanks{Based on
observations made with the William Herschel Telescope operated on the
island of La Palma by the Isaac Newton Group in the Spanish
Observatorio del Roque de los Muchachos of the Instituto de
Astrof\'{i}sica de Canarias.}
}

\author[D. Nestor et al.] 
             {Daniel B. Nestor$^1$, Max Pettini$^1$, Paul C. Hewett$^1$, Sandhya Rao$^2$, and Vivienne Wild$^3$\\
             $^1$Institute of Astronomy, University of Cambridge, 
             Madingley Road, Cambridge CB3 0HA, UK\\
             $^2$Department of Physics and Astronomy, University of Pittsburgh, Pittsburgh, PA 15260, USA\\
             $^3$Max-Planck-Institut f$\ddot{u}$r Astrophysik, 85748 Garching, Germany\\
         } 

\date{Accepted ---; Received ---; in original form ---}
\pagerange{\pageref{firstpage}--\pageref{lastpage}} 
\pubyear{2008}

\maketitle

\label{firstpage}

\begin{abstract}
We present observations of \caii, \znii\, and \crii\ absorption lines in 
16 damped Lyman alpha systems (DLA) and six subDLAs at redshifts
$0.6 < z_{\rm abs} < 1.3$, obtained for the dual purposes of: (i)
clarifying the relationship between DLAs and 
absorption systems selected from their strong \caii\ lines,
and (ii) increasing the still limited sample
of Zn and Cr abundance determinations in this redshift range.
We find only partial overlap between current samples of intermediate
redshift DLAs (which are drawn from magnitude limited surveys) and
strong \caii\ absorbers: approximately 25 per cent of known DLAs at these
redshifts have an associated Ca\,{\sc ii}\,$\lambda 3935$ line with a
rest-frame equivalent width greater than 0.35\,\AA, the threshold of
the Sloan Digital Sky Survey sample assembled by Wild and her
collaborators.  The lack of the strongest \caii\ systems (with
equivalent widths greater than 0.5\,\AA) is consistent with these authors'
conclusion that such absorbers are often missed in current DLA surveys
because they redden and dim the light of the background QSOs.  

We rule out the suggestion that strong \caii\ absorption
is associated exclusively with the highest column density DLAs.
Furthermore, we find no correlation between the strength of the \caii\ lines 
and either the metallicity or degree of depletion of refractory elements, 
although the strongest \caii\ absorber in our sample is also
the most metal-rich DLA yet discovered, with [Zn/H] $\simeq$ solar.  
We conclude that a complex mix of parameters must determine the strengths
of the \caii\ lines, including the density of particles and ultraviolet photons
in the interstellar media of the galaxies hosting the DLAs.  We find
tentative evidence (given the small size of our sample) that strong
\caii\ systems may preferentially sample regions of high gas density, perhaps
akin to the DLAs exhibiting molecular hydrogen absorption at redshifts
$z > 2$.  If this connection is confirmed, strong \caii\ absorbers
would trace possibly metal-rich, H$_2$-bearing columns of cool, dense
gas at distances up to tens of kpc from normal galaxies.

\end{abstract}


\section{Introduction}
\label{Sec:Int} 
Quasar absorption lines are a powerful tool for the study of
gaseous galactic structures over cosmic time, providing clues
to the nature and evolution of galaxies and galactic halos.  The
large columns of neutral hydrogen gas traced by damped Lyman alpha
(DLA) and subDLA systems (conventionally
defined to have \hi\ column densities
$N$(H\,{\sc i}) $\ge 10^{20.3}$\,\acm\ and 
$10^{19} \le N$(H\,{\sc i})\,$ < 10^{20.3}$\,\acm\
respectively) are of particular interest (Rao 2005; Wolfe, Gawiser, 
\& Prochaska 2005), 
as they account for most of the neutral gas in the 
Universe and trace galaxies which are often difficult
to detect directly.  
Thus, appropriate to their importance,
much effort has been expended to understand 
the nature and incidence of
the gas traced by DLAs and subDLAs.

While DLAs do contain metals (Wolfe et al.\ 1986;
Meyer, Welty, \& York 1989; Pettini, Boksenberg, \& Hunstead 1990) 
and ionised gas (Fox et al. 2007a), 
they are characterized by low metallicities
(Kulkarni et al. 2005; Akerman et al. 2005; Prochaska et al. 2007) and
large neutral fractions (Viegas 1995; Vladilo et al. 2001).  
Little, if any, evolution is
detected in the co-moving mass density of \hi, 
$\Omega_{\rm H\,I}$, from $z \simeq 4$ to $z \approx 0.5$ 
(Rao, Turnshek,  \& Nestor, 2006;  Lah et al.\ 2007;  
although see Prochaska, Herbert-Fort, \& Wolfe 2005)
and of metals (Kulkarni et al. 2005).
This is perhaps surprising, considering the putative role of 
DLAs as the reservoir of fuel for star formation.
In fact, recent work (Wolfe \& Chen 2006; Wild, Hewett, 
\& Pettini 2007)
has shown that \textit{in situ} star formation in  DLAs
is significantly less than expected on the basis of the present-day
Kennicutt-Schmidt law (Kennicutt 1998),
considering their large \hi\ surface densities.
One possible explanation for the apparent lack of evolution in 
$\Omega_{\rm H\,I}$ is that a significant fraction of
DLAs with high star formation rates or metallicities  are missed from 
traditional magnitude limited optical QSO surveys due to chromatic 
extinction from large columns of dust. This possibility, however, is not 
supported by surveys of DLAs in radio-selected QSOs (Ellison et al. 2001;
Jorgenson et al. 2006).

SubDLAs, on the other hand, contribute a smaller fraction 
of the total \hi\ in the Universe.  
However, they possess larger fractions of ionised gas than
DLAs, and thus the total amount of hydrogen (\hi\ plus \hii) in
subDLAs may be comparable to that of DLAs (Fox et al. 2007b).  
Furthermore, subDLAs appear to have higher metallicities than 
DLAs (Kulkarni et al. 2007)
and thus may be a large repository of the metals 
produced at high redshifts (see Pettini 2006).  
While DLAs and subDLAs both contain large amounts of hydrogen, 
they are likely to arise in different regions 
within galaxies and to trace different environments and
properties.

Strong metal lines of Mg\,{\sc ii}\,$\lambda\lambda 2796, 2804$ and 
\feii\,$\lambda\lambda 2587, 2600$ are ubiquitous in (sub)DLA systems.  
Their large oscillator strengths and near-ultraviolet (UV)
rest wavelengths make them ideal for tracing large
columns of neutral gas at redshifts $z \la 1.7$, 
where the Ly$\alpha$
line falls below the atmospheric cut-off at 3200\,\AA\ 
and therefore requires observations from space.
In this context, systems selected via
strong Ca\,{\sc ii}\,$\lambda\lambda 3935, 3970$
absorption have recently garnered attention (Wild \& Hewett 2005).
Extensive literature exists on the Ca\,{\sc ii} 
doublet in the interstellar media of the Milky Way and
nearby galaxies, since these lines can be studied
from the ground even at redshift $z=0$
(e.g. Marshall \& Hobbs 1972; 
Welty, Morton, \& Hobbs 1996 and
references therein).
This body of work has shown that in local interstellar
environments Ca is a strongly refractory element,
exhibiting some of the highest depletion factors
(Savage \& Sembach 1996). Furthermore,
since the ionisation potential of Ca\,{\sc ii}
is lower than that of hydrogen, Ca\,{\sc ii} is not
the dominant ionisation stage of Ca in H\,{\sc i}
regions, unlike Mg\,{\sc ii} and Fe\,{\sc ii}.
Rather, the balance between Ca\,{\sc ii} and Ca\,{\sc iii}
depends on the details of the physical conditions 
in the gas, primarily on temperature and on the 
densities of particles and far-UV photons.
Therefore, while all DLAs 
exhibit strong \mgii\ and \feii\ lines, the same
is not true of \caii.  

Wild, Hewett, \& Pettini (2006; WHP06) recently 
conducted a survey for strong \caii\ systems, 
with $\lambda 3935$ rest equivalent width
\wca\,$\geq 0.35$\,\AA, in QSO spectra from the
Sloan Digital Sky Survey (SDSS).  
Based on the strengths of Zn\,{\sc ii}\ and Cr\,{\sc ii}\ 
lines, WHP06 concluded that, on average, strong
Ca\,{\sc ii} absorbers have \nhi\ values above the
DLA limit.  
They also showed that these absorbers are, on average,  
dustier than typical DLAs, introducing a reddening
of $\langle E(B-V)\rangle \ga 0.1$\,magnitudes 
in the spectra of the background QSOs,
compared to  $\langle E(B-V)\rangle \la 0.02$ 
for the general population of SDSS DLAs 
(WHP06; Vladilo, Prochaska, \& Wolfe 2008).
This level of obscuration of the 
background QSOs is significant, as it 
causes flux-limited and/or colour-selected surveys to 
underestimate the numbers of strong Ca\,{\sc ii} systems.  
If \caii-strong DLAs are being missed, 
and if they are preferentially metal-rich relative
to DLAs with weaker \caii\ absorption, 
then this class of absorption
system may contain a previously overlooked 
repository of both neutral gas and metals.
In any case, if \caii\ systems select a population of
absorbers with unique properties, they could hold the key 
to a better understanding of the detailed physical
properties of DLAs and subDLAs.

Since the survey by WHP06, efforts have been devoted to better
understanding the link between strong Ca\,{\sc ii} systems 
and galaxies. Using $K$-band imaging, Hewett \& Wild (2007)
reported an excess of luminous galaxies close to the sightlines
to 30 QSOs with intervening Ca\,{\sc ii} systems at redshifts
$0.7 < z < 1.2$.  
The galaxies appear to exhibit a luminosity-dependent
cross section for \caii\ absorption and have a mean impact parameter 
of $\approx 25$\,kpc.
Together with the incidence of absorbers
from WHP06, this result implies a halo volume filling factor of 
$\sim 10$ per cent out to $\approx 35$\,kpc from the
absorbing galaxies.
At lower redshifts, Zych et al. (2007) were able to associate
luminous, metal-rich, star-forming spiral galaxies 
with four out of five strong Ca\,{\sc ii} absorbers at $z < 0.5$.

Although these studies represent progress in uncovering 
the nature of \caii\ absorbers, ignorance of \nhi\  
in such systems limits the direct insight they provide into DLAs and 
subDLAs in general.  
While it seems very likely that \caii\ absorbers select large columns of
neutral gas, direct measurements of \nhi\ are currently
available for very few \caii\ systems.
The reason is simple: by the time the SDSS data highlighted the 
potential importance of absorption systems selected via
Ca\,{\sc ii}, no space-borne UV spectrograph was in operation to
measure the associated column densities of H\,{\sc i}.
This unfortunate state of affairs is about to change with the forthcoming
installation of the Cosmic Origins Spectrograph (COS) on the 
refurbished \textit{Hubble Space Telescope} (\textit{HST}).
In the meantime, the most immediate approach to this problem
is to perform the complementary experiment of measuring
the strength of \caii\ absorption in known DLAs at redshifts
$z \simlt 1.3$, where the \caii\ doublet lines are still accessible
with optical spectrographs.

The samples of confirmed DLAs at these intermediate
redshifts have increased considerably in recent years
thanks to dedicated \textit{HST} surveys
(see Rao, Turnshek \& Nestor 2006; RTN06).
In this paper, we present new observations of QSOs
from the RTN06 compilation designed to: \textit{(i)}
measure the equivalent widths of the
Ca\,{\sc ii}\,$\lambda \lambda 3935, 3970$ doublet
lines in order to assess the overlap of known DLAs 
(and a few subDLAs) with the
strong Ca\,{\sc ii} absorber sample of WHP06; and \textit{(ii)}
simultaneously determine the metallicities and dust content
of intermediate redshift DLAs via observations
of the associated Zn\,{\sc ii} and Cr\,{\sc ii} absorption lines
(Pettini et al. 1990), thereby adding to the still somewhat
limited statistics of these measures 
at $z < 1.5$ (Akerman et al. 2005; Kulkarni et al. 2007).
In Section~\ref{Sec:Obs}
we describe our observations and equivalent width measurements,
while Section~\ref{sec:abundances} deals with the
derivation of column densities and element abundances.
The main results on the \caii\ properties of known
DLAs at intermediate redshifts are presented in Section~\ref{sec:caii_in_dlas}.
We summarise our principal findings and conclusions
in Section~\ref{Sec:Sum}.

\section{Observations and Data Reduction}
\label{Sec:Obs} 

\begin{table*}
\centering
\begin{minipage}[c]{0.95\textwidth}
\caption{\textsc{DLAs observed, equivalent widths, column densities, and abundances}}
\begin{tabular}{@{}lccccccccc}
\hline
   \multicolumn{1}{c}{QSO}
& \multicolumn{1}{c}{$z_{\rm em}$}
& \multicolumn{1}{c}{$z_{\rm abs}$}
& \multicolumn{1}{c}{\wmg}
& \multicolumn{1}{c}{\wca}
& \multicolumn{1}{c}{$\log N$(H\,{\sc i}) }
& \multicolumn{1}{c}{$\log N$(Zn\,{\sc ii}) }
& \multicolumn{1}{c}{$\log N$(Cr\,{\sc ii}) }
& \multicolumn{1}{c}{[Zn/H]}
& \multicolumn{1}{c}{[Cr/Zn]}\\
   \multicolumn{1}{c}{}
& \multicolumn{1}{c}{}
& \multicolumn{1}{c}{}
& \multicolumn{1}{c}{(\AA)}
& \multicolumn{1}{c}{(\AA)}
& \multicolumn{1}{c}{(cm$^{-2}$)}
& \multicolumn{1}{c}{(cm$^{-2}$)}
& \multicolumn{1}{c}{(cm$^{-2}$)}
& \multicolumn{1}{c}{}
& \multicolumn{1}{c}{}\\
   \multicolumn{1}{c}{(1)}
& \multicolumn{1}{c}{(2)}
& \multicolumn{1}{c}{(3)}
& \multicolumn{1}{c}{(4)}
& \multicolumn{1}{c}{(5)}
& \multicolumn{1}{c}{(6)}
& \multicolumn{1}{c}{(7)}
& \multicolumn{1}{c}{(8)}
& \multicolumn{1}{c}{(9)}
& \multicolumn{1}{c}{(10}\\
\hline
0139$-$0023   & 1.384  &  0.6825  & 1.24  & $0.24 \pm 0.03$   & 20.60$^{+0.05}_{-0.12}$   & $<13.15$               & $<13.66$                 & $ <-0.08 $              &  \ldots \\
0253$+$0107  & 1.035  &  0.6316  & 2.57  & $0.46 \pm 0.04$     & 20.78$^{+0.12}_{-0.08}$    & $<13.06$               & $<13.69$                 & $ <-0.35 $              & \ldots  \\
0256$+$0110  & 1.349  &  0.7252  & 3.10  &$ 0.39 \pm 0.04$     & 20.70$^{+0.11}_{-0.22}$    & $ 13.20 \pm 0.09^{\rm a}$ & $ 12.72 \pm 0.68^{\rm a}$   & $ -0.13 \pm 0.24^{\rm a}$  & $-1.50 \pm 0.68^{\rm a}$\\
0449$-$168    & 2.679  &  1.0072  & 2.14  &$<0.04$             & 20.98$^{+0.06}_{-0.07}$    & ($12.62 \pm 0.07$)$^{\rm \dag}$ &  ($13.47 \pm 0.02$)$^{\rm \dag}$ & $-0.99 \pm 0.10$    & $-0.17 \pm 0.07$  \\
0454$+$039   & 1.343  &  0.8591  & 1.45  &$ 0.063 \pm 0.008$  & 20.67$\pm0.03$    & $ 12.42 \pm 0.06$ & $ 13.44 \pm 0.02$   & $ -0.88 \pm 0.06$  & $+0.00 \pm 0.06$ \\
1007$+$0042  & 1.681 &  1.0373  & 2.98  & $<0.08$                   & 21.15$^{+0.15}_{-0.24}$    & $ 13.27 \pm 0.04$ & $ 13.55 \pm 0.08$   & $ -0.51 \pm 0.24$  & $-0.74 \pm 0.09$ \\
1009$-$0026  & 1.244  &  0.8426  & 0.71  & $<0.05$                   & 20.20$^{+0.05}_{-0.06}$    & $<12.36^{\rm b}$              & $<13.11^{\rm b} $                & $ < -0.47^{\rm b} $              & \ldots\\
                     &	         &  0.8860  & 1.90  & $ 0.07 \pm 0.02$     & 19.48$^{+0.01}_{-0.08}$    & $ 12.43 \pm 0.14^{\rm b}$ & $ <13.10^{\rm b} $               & $ 0.32 \pm 0.15^{\rm b} $  & $< -0.35^{\rm b} $\\
1010$+$0003 & 1.399  &  1.2642  & 1.12  & $ 0.26 \pm 0.02$     & 21.52$^{+0.06}_{-0.07}$    & $ 13.01 \pm 0.02$ & $ 13.69 \pm 0.02$   & $ -1.14 \pm 0.07$  & $-0.34 \pm 0.03$ \\
1107$+$0048 & 1.392  &  0.7402  & 2.95  & $ 0.34 \pm 0.01$   & 21.00$^{+0.02}_{-0.05}$     & $ 13.12 \pm 0.04$ & $ 13.75 \pm 0.04$   & $ -0.52 \pm 0.07$ & $-0.39 \pm 0.06$\\
1107$+$0003 & 1.740  &  0.9542  & 1.36  & $ 0.26 \pm 0.02$     & 20.26$^{+0.09}_{-0.14}$    & $<12.38^{\rm c}$                & $<13.18^{\rm c}	  $               & $ <-0.51^{\rm c} $              & \ldots \\
1220$-$0040  & 1.411 &  0.9746   & 1.95  & $<0.08 $                  & 20.20$^{+0.05}_{-0.09}$     & $<12.55^{\rm d} $              & $<13.11^{\rm d} $                & $ <-0.28^{\rm d} $              & \ldots  \\
1224$+$0037 & 1.482 &  1.2346   & 1.09  & $<0.16 $                  & 20.88$^{+0.04}_{-0.06}$     & $<12.61^{\rm b} $              & $<13.30^{\rm b} $                & $ <-0.90^{\rm b} $              & \ldots \\
                     &          &  1.2665   & 2.09  & $<0.15 $                 & 20.00$^{+0.08}_{-0.05}$      & $<12.62^{\rm b} $              & $<13.03^{\rm b} $                & $ <-0.01^{\rm b} $              & \ldots\\
1225$+$0035 & 1.226  & 0.7728   & 1.74  & $ 0.36 \pm 0.03$   & 21.38$^{+0.11}_{-0.12}$     & $ 13.23 \pm 0.07$ & $ 13.81 \pm 0.06$   & $ -0.78 \pm 0.14$  & $-0.44 \pm 0.09$\\
1323$-$0021  & 1.390 &  0.7156   & 2.23  & $ 0.82 \pm 0.02$    & 20.54$\pm0.15$      & $ 13.29 \pm 0.21^{\rm a}$ & $ 13.70 \pm 0.18$   & $ +0.12 \pm 0.26$ & $-0.61 \pm 0.28$\\
1727$+$5302 & 1.444 &  0.9445   & 2.83  & $ 0.29 \pm 0.03$    & 21.16$^{+0.04}_{-0.05}$      & $ 13.23 \pm 0.03$ & $ 13.80 \pm 0.03$    & $ -0.56 \pm 0.06$  & $-0.45 \pm 0.04$\\
                     &          &  1.0306   & 0.92  & $ 0.16 \pm 0.02$   & 21.41$\pm0.03$       & $ 12.68 \pm 0.07^{\rm e}$ & $ 13.36 \pm 0.06^{\rm e}$   & $ -1.36 \pm 0.08^{\rm e}$  & $-0.34 \pm 0.09^{\rm e}$\\
1733$+$5533  & 1.072 &  0.9981  & 2.17  & $<0.07 $                 & 20.70$^{+0.04}_{-0.03}$       & $ 12.89 \pm 0.06$ & $ 13.34 \pm 0.07$   & $ -0.44 \pm 0.07$  & $-0.57 \pm 0.09$ \\
2149$+$212   & 1.538 &  0.9111   & 0.72  & $ 0.12 \pm 0.01$   & 20.70$^{+0.08}_{-0.10}$       & $<12.40 $               & $<12.78  $               & $<-0.93  $              & \ldots \\
                    &           &  1.0023   & 2.46  & $<0.11 $                & 19.30$^{+0.02}_{-0.05}$       & $<12.13 $               & $<12.59  $               & $ <+0.20 $             & \ldots \\
2353$-$0028  & 0.765 &  0.6043   & 1.60  & $ 0.17 \pm 0.03$   & 21.54$\pm0.15$       & $ 13.25 \pm 0.29$ & $ 13.40 \pm 0.17$    & $ -0.92 \pm 0.32$ & $-0.86 \pm 0.33$ \\
\hline
\end{tabular}
\\
\dag\ from P\'{e}roux et al. (2008).  Measurements having greater precision are reported by: (a) P\'{e}roux et al. (2006a); 
(b) Meiring et al. (2007); (c) Meiring et al. (2006); (d) Meiring et al. (2008); (e) Khare et al. (2004).\\
\label{table:data}
\end{minipage}
\end{table*}

\begin{table*}
\centering
\begin{minipage}[c]{1.025\textwidth}
\caption{\textsc{Measurements from the literature}}
\begin{tabular}{@{}lcccccccccc}
\hline
   \multicolumn{1}{c}{QSO}
& \multicolumn{1}{c}{$z_{\rm em}$}
& \multicolumn{1}{c}{$z_{\rm abs}$}
& \multicolumn{1}{c}{\wmg}
& \multicolumn{1}{c}{\wca}
& \multicolumn{1}{c}{$\log N$(H\,{\sc i}) }
& \multicolumn{1}{c}{$\log N$(Zn\,{\sc ii}) }
& \multicolumn{1}{c}{$\log N$(Cr\,{\sc ii}) }
& \multicolumn{1}{c}{[Zn/H]}
& \multicolumn{1}{c}{[Cr/Zn]}
& \multicolumn{1}{c}{Ref.$^{\rm b}$}\\
   \multicolumn{1}{c}{}
& \multicolumn{1}{c}{}
& \multicolumn{1}{c}{}
& \multicolumn{1}{c}{(\AA)}
& \multicolumn{1}{c}{(\AA)}
& \multicolumn{1}{c}{(cm$^{-2}$)}
& \multicolumn{1}{c}{(cm$^{-2}$)}
& \multicolumn{1}{c}{(cm$^{-2}$)}
& \multicolumn{1}{c}{}
& \multicolumn{1}{c}{}
& \multicolumn{1}{c}{}\\
   \multicolumn{1}{c}{(1)}
& \multicolumn{1}{c}{(2)}
& \multicolumn{1}{c}{(3)}
& \multicolumn{1}{c}{(4)}
& \multicolumn{1}{c}{(5)}
& \multicolumn{1}{c}{(6)}
& \multicolumn{1}{c}{(7)}
& \multicolumn{1}{c}{(8)}
& \multicolumn{1}{c}{(9)}
& \multicolumn{1}{c}{(10)}
& \multicolumn{1}{c}{(11)}\\
\hline
0515$-$4414      & 1.713   &  1.1508   &  2.34     & $ 0.36 \pm 0.07$      & $19.88\pm0.05$            & $12.18\pm 0.02$   & $12.48\pm0.04$     &  $-0.33\pm 0.05$         & $-0.72\pm0.04$   & 1,2 \\ 
0738$+$313       & 0.630   &  0.0912   &  \ldots     & $ 0.19 \pm 0.01$      & $21.18^{+0.05}_{-0.07}$                   & $<12.662$              & $13.28 \pm 0.22$   &  $<-1.15$               &  $> -0.40$  & 3 \\
0738$+$313       & 0.630   &  0.2210   &  0.61         & $ 0.06 \pm 0.01$      & $20.90^{+0.07}_{-0.08}$             & $<12.825$              & $13.11 \pm 0.24$   &  $<-0.71$               &  $> -0.74$  & 3 \\
1436$-$0051      & 1.275   &  0.7377   &  1.11         & $ 0.40 \pm 0.02$      & $20.08^{+0.10}_{-0.12}$ & $12.67\pm 0.05$   & $<12.71 $                &  $-0.04 \pm 0.12$  & $< -0.98$   & 4\\
2328$+$0022     & 1.309   &  0.652     &  1.90     & $ 0.24 \pm 0.02$      & $20.32^{+0.06}_{-0.07}$  & $12.43 \pm 0.15$   & $13.35 \pm 0.19$      &  $-0.52 \pm 0.22$                & $-0.10 \pm 0.24$  & 5 \\
2331$+$0038     & 1.486   &  1.1414   &  2.53         & $0.26 \pm 0.02$       & $20.00^{+0.04}_{-0.05}$  & $12.22 \pm 0.09$   & $<12.37$                    & $ -0.41 \pm 0.12$               & $< -0.87$  & 6 \\
2335$+$1501     & 0.790   &  0.6798   &  0.97         & $0.19 \pm 0.03$       & $19.70^{+0.30}_{-0.30}$  & $12.37 \pm 0.04$   & $ 12.89 \pm 0.10 $          & $ +0.04 \pm 0.30$               & $-0.50 \pm 0.11$  & 7 \\
\hline
\end{tabular}
\\
$^{\rm b}${References---1: de la Varga et al. (2000);
2: Quast et al. (2008); 
3: Meiring et al. (2006);
4: Meiring et al. (2008);
5: P\'{e}roux et al. (2006a);
6: Meiring et al. (2007);
7: P\'{e}roux et al. (2008).
}\\
\label{table:others}
\end{minipage}
\end{table*}

\begin{figure*}
   \centering
  \includegraphics[angle=0,width=130mm]{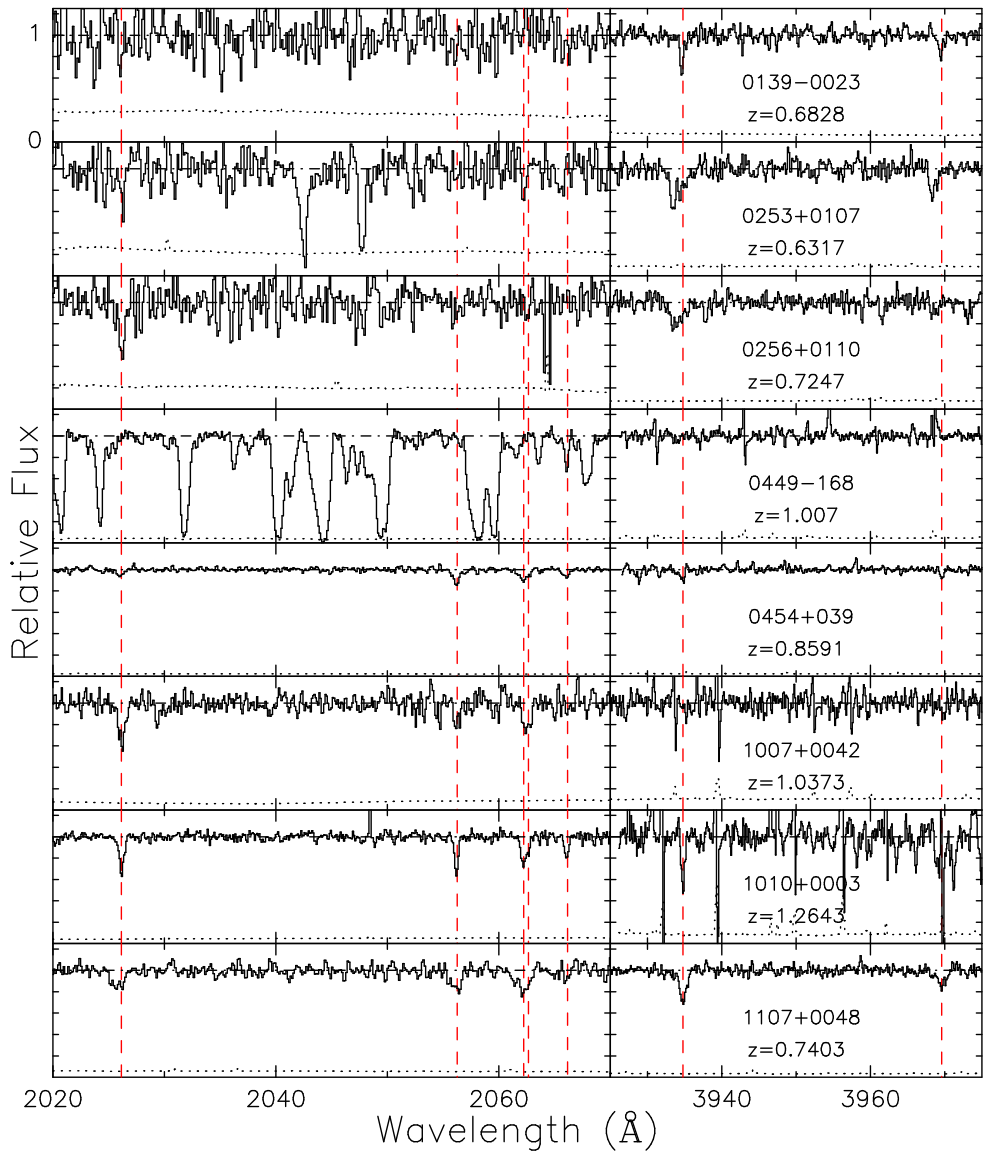}
  \caption{The \znii-\crii\ (left) and \caii\ (right) regions
of the spectra of the 16 DLAs observed.
The spectra have been normalised to the underlying QSO
continuum and reduced to the rest frame of the DLA.
In each panel, the horizontal dash-dot line 
indicates the continuum level, while the dotted line
near the bottom of each plot shows the $1 \sigma$ error
spectrum.
Vertical long-dash lines are drawn at the wavelengths of
the Zn\,{\sc ii}$\lambda \lambda 2026, 2063$ doublet, 
Cr\,{\sc ii}\,$\lambda \lambda 2056, 2062, 2066$ triplet,
and Ca\,{\sc ii}\,$\lambda \lambda 3935, 3970$ doublet. 
}
  \label{Fig:DLAspecs}
\end{figure*}

\subsection{Sample Selection}

We selected QSOs with known DLAs and subDLAs at
$0.6 < z_{\rm abs} < 1.3$ from the compilation by RTN06.  
Targets were chosen on the basis of  absorption redshift 
(so as to allow coverage of the absorption features of
interest), and accessibility during the scheduled observing runs.
For each target, we first examined the SDSS spectrum at the wavelengths
of the expected \caii\ absorption.  However, those spectra
have insufficient signal-to-noise ratios (S/N) and resolution 
to either unambiguously detect or put meaningful limits 
on \caii\ absorption strength.

It is important to remember here that the sample assembled by RTN06
was constructed from a \textit{targeted} Ly$\alpha$ survey 
of known \mgii\ absorbers. 
Therefore, as with any absorption line study, the manner in which
systems are selected needs to be considered when interpreting the results
of our survey.  
For example, stronger \mgii\ systems are exponentially less
common than weaker systems (Nestor, Turnshek, \& Rao 2005), but
more likely to be DLAs (RTN06).  Combining these two trends,
one can demonstrate that the unbiased median \wmg\ for DLAs 
is $\simeq 1.35$\,\AA, compared to 1.9\,\AA\ in our sample. 
Evidently, our sample is somewhat biased against DLAs with 
weaker \mgii\ absorption, compared to a blind survey.

Celestial coordinates and magnitudes for each 
quasar in our sample are given in Appendix A.

\setcounter{figure}{0}
\begin{figure*}
   \centering
  \includegraphics[angle=0,width=130mm]{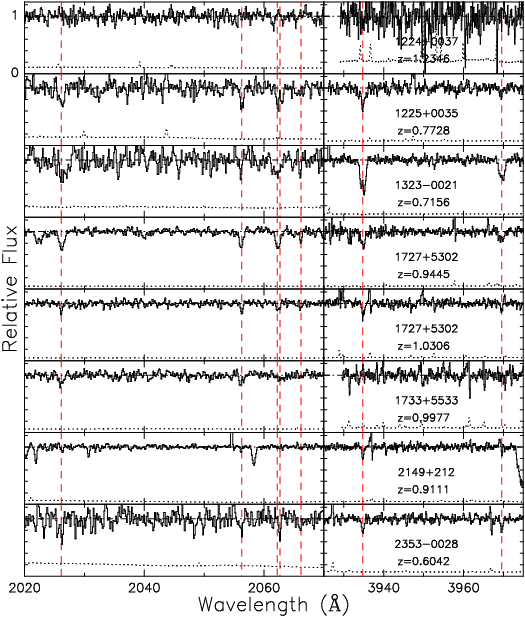}
  \caption{\textit{(cont'd)} 
}
\end{figure*}

\subsection{Observations}

Observations were carried out with 
the ISIS double-beam spectrograph 
on the 4.2\,m William Herschel telescope (WHT) 
at the Roque de Los Muchachos
Observatory, La Palma,  in two observing runs of three
nights each, in 2006 April and 2006 November respectively.
Atmospheric conditions varied from partly cloudy to clear,
with seeing in the range 0.6--2.0\,arcsec.
In total we secured spectra of 18 QSOs covering absorption lines
of 16 DLAs and six subDLAs, as detailed in Table~\ref{table:data}.

Each observation consisted of a series of simultaneous
exposures in the blue and red arms of ISIS; individual
exposures typically lasted 1800\,s, and the total
exposure times in each arm ranged from 1 to 5.5 hours per QSO
(2--3 hours in most cases).
The blue arm was configured so as to cover the region 
encompassing the \znii\ and \crii\ multiplets,
while the red arm was centred on the redshifted
\caii\ doublet.  
With 1200\,grooves~mm$^{-1}$ gratings
and a 0.8\,arcsec entrance slit, ISIS
delivered a spectral resolution of 0.6--0.7\,\AA\ full width
at half maximum (FWHM), 
sampled with three pixels of the EEV (blue arm) and Marconi (red arm) CCDs.
Bright B-type stars were
observed each night to provide templates for the
removal of telluric absorption lines.
The emission line spectra of hollow-cathode Cu-Ar and Cu-Ne 
lamps were recorded at hourly intervals to be used for subsequent
wavelength calibration. 
Spectrophotometric standards from the 
compilation by Massey et al. (1988) allowed for flux calibration
of the QSO spectra.

\subsection{Data Reduction and Measurements}
\label{sec:data_redux}

The reduction of the two-dimensional images into
co-added one-dimensional spectra followed standard
procedures. 
We used {\sc iraf}\footnote{{\sc iraf} is distributed by the National Optical
Astronomy Observatories, which are operated by the Association of
Universities for Research in Astronomy, Inc.  under cooperative
agreement with the National Science Foundation.}
scripts to carry out the 
initial stages of bias subtraction,
flat-fielding, wavelength calibration,
background subtraction, and spectral extraction.
The individual spectra of each QSO were co-added
with weights proportional to the S/N
and then binned
onto a linear, vacuum heliocentric wavelength
scale with approximately three bins per resolution
element. The final steps in the data reduction
involved correction for telluric absorption
(if it occurred close to 
absorption lines of interest), 
and normalisation of each spectrum to the local QSO continuum.

\begin{figure*}
   \centering
  \includegraphics[angle=0,width=130mm]{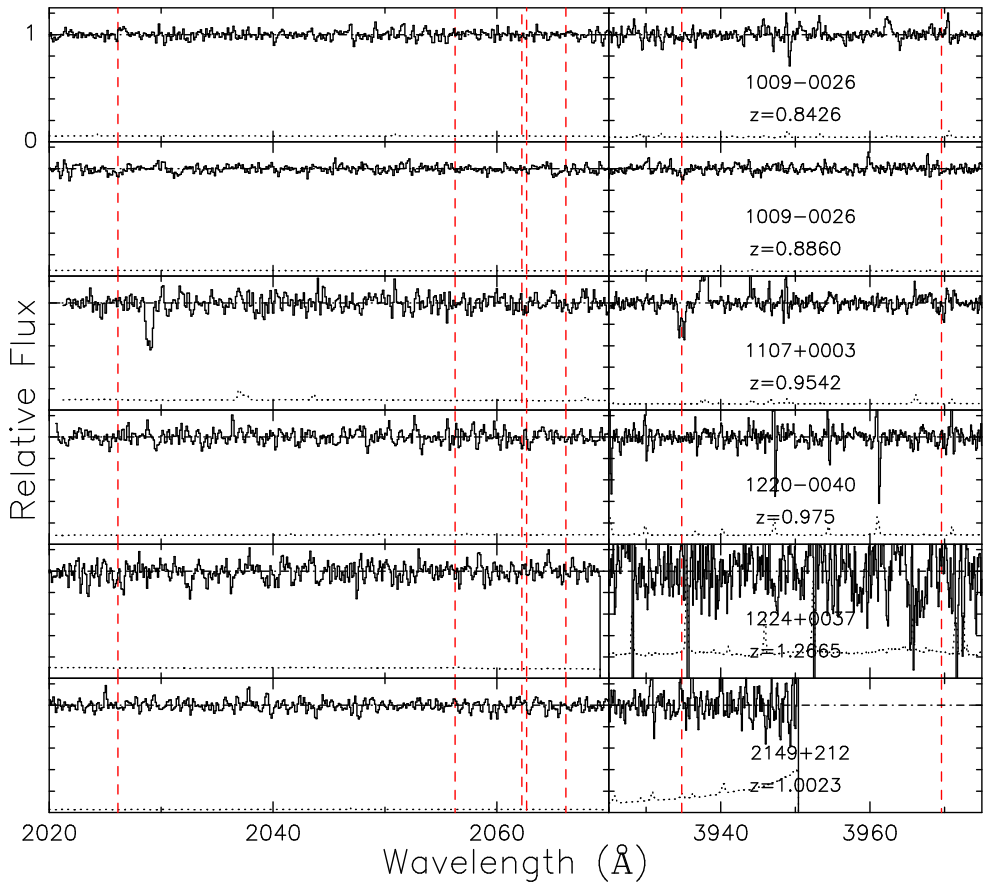}
  \caption{As in Figure~\ref{Fig:DLAspecs}, but for the six
  subDLAs observed.
}
  \label{Fig:sDLAspecs}
\end{figure*}

In Figures~\ref{Fig:DLAspecs} and \ref{Fig:sDLAspecs}
we have reproduced normalised portions of the spectra
encompassing the Zn\,{\sc ii}, Cr\,{\sc ii} and Ca\,{\sc ii}
absorption lines in, respectively,  the 16 DLAs and six subDLAs,
ordered in increasing Right Ascension.
As can be appreciated from inspection of these two figures,
the S/N achieved differs significantly from QSO to QSO, 
depending on the variable sky transparency, seeing, 
QSO magnitude
and redshifted wavelengths of the features of interest.
In every case, however, the combination of S/N and
spectral resolution is sufficient to detect a 
Ca\,{\sc ii}\,$\lambda 3935$ absorption line
with a rest equivalent width \wca\,$= 0.35$\,\AA\
(the threshold of the SDSS survey by WHP06)
at the $\geq 5 \sigma$ significance level.

Values of the \wca\ are listed in column (5) of 
Table~\ref{table:data}; $3 \sigma$ upper limits
are given for cases where no Ca\,{\sc ii} absorption 
could be identified. 
We detected the stronger member of the Ca\,{\sc ii} doublet,
$\lambda 3935$, in 12 out of the 16 DLAs targeted;
however, only four of the DLAs meet the selection
criterion \wca\,$\geq 0.35$\,\AA\ of the survey
by WHP06. We return to this point in Section \ref{sec:caii_in_dlas}
below.
For reference, we also give in column (4) of Table~\ref{table:data}
the rest-frame equivalent width of the stronger member of the
Mg\,{\sc ii} doublet, \wmg, from the work by RTN06.

While this study represents the first systematic study of Ca\,{\sc ii}
absorption in (sub)DLAs, UV absorption lines in many of the systems 
in our sample have previously been studied by other authors.  However,
for five of the 16 DLAs and one of the six subDLAs,
no previous measurements of the abundances of Zn and Cr have been 
reported.  For four other systems, improved data for Zn and Cr are
reported, two of which are significant and two which are
marginal improvements over published results.
Cases where the previous measurements are of superior precision to ours
are indicated by notes in columns (7) and (8) of Table~\ref{table:data}.

As well as considering the individual spectra, we also
constructed stacked spectra by adding together the
normalised spectra of each DLA in the
Zn\,{\sc ii}-Cr\,{\sc ii} and Ca\,{\sc ii} regions
(in the DLA rest frame), following the same procedure as WHP06.
The resulting average spectrum, which includes DLAs in which 
the Ca\,{\sc ii} lines are below our detection limit, 
is useful for comparative studies, as discussed below.
A second average spectrum was assembled by adding
together the data for each subDLA.
The borderline DLA towards 1107+0003 
($\log N$(H\,{\sc i})\,$ = 20.26 \pm 0.1$) was included
in the DLA stack; the inclusion/exclusion of this system in the
DLA/subDLA stack has little effect on the results.
The \znii-\crii\ region of the DLA towards 0449$-$168 
was excluded from all stacks, as it falls within the
Ly$\alpha$ forest (see Figure~\ref{Fig:DLAspecs}).

\section{Column Densities and Element Abundances}
\label{sec:abundances}

Column densities of Zn\,{\sc ii} and Cr\,{\sc ii}
were measured by fitting the observed absorption lines
with theoretical profiles generated with the software
package {\sc vpfit}\footnote{{\sc vpfit} is available from
http://www.ast.cam.ac.uk/\textasciitilde rfc/vpfit.html} 
as described, for example, by Rix et al. (2007).
Briefly, {\sc vpfit} uses $\chi^2$ minimisation to 
deduce the values of redshift $z$, column density $N$ (cm$^{-2}$), 
and Doppler parameter $b$ (km~s$^{-1}$)
that best reproduce the observed absorption line profiles.
{\sc vpfit} takes into account the 
instrumental broadening function
in its $\chi^2$ minimisation and error evaluation.
Laboratory (vacuum) wavelengths and $f$-values of the
transitions analysed were taken from the compilation by Morton (2003)
with subsequent updates by Jenkins \& Tripp (2006).
In most cases, the resolution of our spectra 
(FWHM\,$\approx 30$--60\,km~s$^{-1}$)
is  insufficient to resolve the intrinsic 
velocity structure of the gas.  
However, as we are dealing with 
weak absorption lines, we do not expect
line saturation to be a problem
(see the discussion by Pettini et al. 1990);
the measured doublet (Zn\,{\sc ii}) and triplet
(Cr\,{\sc ii}) ratios confirm that this is indeed the case
for the systems analysed here.

Values of $N$(Zn\,{\sc ii}) and $N$(Cr\,{\sc ii}) are listed
in Table~\ref{table:data}, together with the errors estimated
by {\sc vpfit} on the basis of the S/N ratios of the spectra.
Since these species are the dominant
ionisation stages of Zn and Cr in H\,{\sc i} regions
(e.g. Morton et al. 1973; Viegas 1995), the corresponding
element abundances can be inferred directly from the
ratios $N$(Zn\,{\sc ii})/$N$(H\,{\sc i}) and 
$N$(Cr\,{\sc ii})/$N$(H\,{\sc i}). 
Values of $N$(H\,{\sc i}) for the systems observed 
are reproduced in column (6) of Table~\ref{table:data} 
from the compilation by RTN06.
Reference to the solar values of Zn/H and Cr/Zn
($\log {\rm (Zn/H)}_{\odot}  = -7.37$ and
$\log {\rm (Cr/Zn)}_{\odot}  = +1.02$; 
Lodders 2003), then gives
the quantities [Zn/H] and [Cr/Zn] in the usual
definition [X/Y]\,$\equiv \log {\rm (X/Y)}_{\rm DLA} - \log {\rm (X/Y)}_{\odot}$.
The former is often adopted as a measure of 
the metallicity, while the latter provides an indication
of the degree of dust depletion of refractory
elements, for the reasons discussed by Pettini et al. (1990).
Values of [Zn/H] and [Cr/Zn] for the systems observed 
are given in the last two columns of Table~\ref{table:data};
the errors quoted were obtained by combining in quadrature
the errors on the column densities of the respective ions.

\begin{figure}
   \centering
 {\hspace*{-0.25cm} \includegraphics[angle=0,width=80mm]{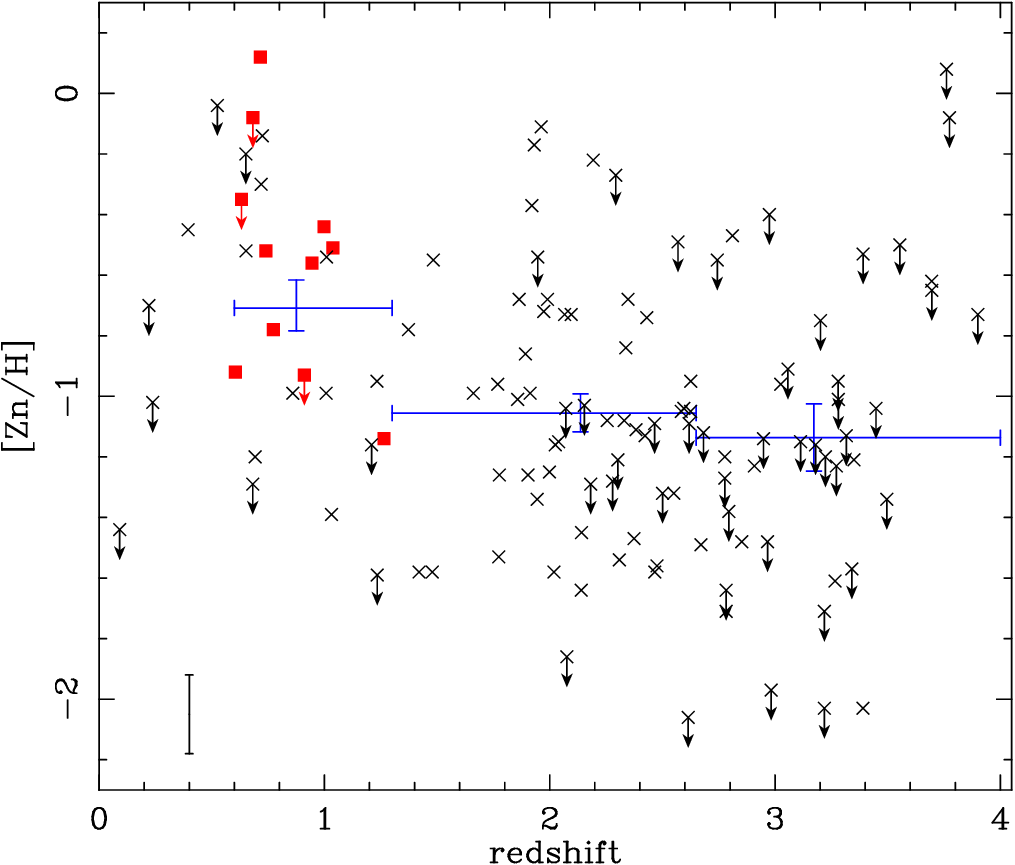}}
  \caption{Measurements of [Zn/H] in DLAs at $0 < z_{\rm abs} < 4$
from the compilation by Kulkarni et al. (2007) and from the present
work.  Except for the cases in which there exists a superior
measurement in the literature, the DLAs from the present study are
shown as red squares. Upper-limits are indicated with 
downward pointing arrows.  Also shown are the mean values of  column
density-weighted metallicity [$\langle {\rm Zn/H} \rangle]$ (see text
for definition) in three redshift intervals, together with the $\pm 1
\sigma$ ranges calculated via bootstrap techniques. As discussed in
the text,  the value of [$\langle {\rm Zn/H} \rangle]$ at $0.6 <
z_{\rm abs} \leq 1.3$ may be biased too high by about 0.1\,dex,
because these DLAs were primarily selected via their strong Mg\,{\sc
ii} absorption.  The vertical bar in the lower left-hand corner gives
an indication of the typical error in [Zn/H].  }
  \label{Fig:znhz}
\end{figure}

As noted above, measurements of Zn and Cr abundances have been previously
reported in the literature for many of the absorbers in our sample.
Out of the ten DLAs in common with previous work for which we have
estimates of [Zn/H] and [Cr/Zn], eight are in very good agreement with
published values. 
The exceptions are the DLAs in the spectra of 
the QSOs 1225+0035 and 1733+5533 
(at $z_{\rm abs} = 0.7728$ and 0.9981 respectively)
where the values of [Zn/H] implied from
our detections of the Zn\,{\sc ii} doublet lines
(see Figure~\ref{Fig:DLAspecs}) are 
actually higher than the upper limits estimated by
Meiring et al. (2006). However, these authors' data
are of rather poor S/N at the relevant wavelengths.

\subsection{Metallicity Evolution of DLAs}

The most recent compilation of [Zn/H] measurements
in DLAs by Kulkarni et al. (2007) includes 18 systems
in the redshift range $0.6 < z < 1.3$ 
(corresponding to a time interval amounting 
to nearly one quarter of the current age of the universe).
Thus, the five new (and two improved) measurements
reported here constitute a significant addition to
the DLA statistics at these epochs. 
In Figure~\ref{Fig:znhz} we plot the full sample of
[Zn/H] measurements as a function of redshift.  
The new/improved data reported here, shown with solid squares,
confirm the wide range of metallicities exhibited 
by DLAs \emph{at all redshifts}: as pointed out by
Pettini et al. (1994, 1997b), at any one epoch DLA
metallicities can span two orders of magnitude, from
nearly solar to $\sim 1/100$ of solar. 

Also shown in Figure~\ref{Fig:znhz} are the mean values
of the column density-weighted metallicity,
[$\langle {\rm Zn/H} \rangle$]
(where $\langle {\rm Zn/H} \rangle = \sum N$(Zn\,{\sc ii})$/\sum N$(H\,{\sc i})---see
Pettini et al. 1994),
for the redshift range of our sample 
($0.6 < z_{\rm abs}  < 1.3$) and for higher redshift DLAs 
in two bins: $1.3 < z_{\rm abs}  < 2.6$ and $2.6 < z_{\rm abs} < 4$.
As discussed by Kulkarni et al. (2007 and references therein),
there is evidence for mild evolution of the mean DLA
metallicity with
the progress of time, in the sense that the value
[$\langle {\rm Zn/H} \rangle] = -0.71 \pm 0.08$ we measure
at $0.6 < z_{\rm abs}  \leq 1.3$
is a factor of about two to three higher than in the other two redshift
bins: [$\langle {\rm Zn/H} \rangle] = -1.06 \pm 0.06$
at $1.3 < z_{\rm abs}  \leq 2.6$, and
[$\langle {\rm Zn/H} \rangle] = -1.14 \pm 0.11$
at $2.6 < z_{\rm abs}  \leq 4.0$, respectively.

One of the factors which affect this type of comparison,
and yet is generally overlooked, is the fact that DLAs
with $z_{\rm abs} < 1.7$ are normally selected via their
(strong) Mg\,{\sc ii}\,$\lambda 2796$ absorption, while
the selection at higher redshifts is via the damped \lya\
line itself. The two criteria may not be equivalent; in
particular, it now seems well established  
that DLAs with strong Mg\,{\sc ii}
absorption are on the whole more metal-rich 
than the average (Nestor et al. 2003; Murphy et al. 2007). 
To assess the importance of this potential
bias, we use four distributions: \textit{(i)} the distribution
of DLA metallicities as a function of \wmg\ -- 
at redshifts $0.6 < z < 1.3$, we find that
[Zn/H]\,$\propto $\,(\wmg)$^{1.7 \pm 0.1}$ is a good
representation of the 12 measured values in our sample
(Pearson's r $= 0.85 \pm 0.08$, $\chi^2_\nu = 2.7$); 
\textit{(ii)} the DLA fraction as a function of \wmg\
from RTN06; \textit{(iii)} the distribution of values
of \wmg\ in SDSS QSOs from Nestor et al. (2005); and \textit{(iv)}
the distribution of values of \wmg\ in the sample used to
estimate [$\langle {\rm Zn/H} \rangle]$, to arrive at the 
conclusion that the value of 
[$\langle {\rm Zn/H} \rangle]= -0.7 \pm 0.08$ 
at $0.6 < z_{\rm abs}  < 1.3$
is probably biased too high by $\sim 0.1$\, dex.
Although this is not a large factor, it nevertheless
dilutes further the evidence for redshift evolution
of the mean metallicity of DLAs.

\subsection{Depletion of Chromium}
\label{sec:crzn}

\begin{figure} \vspace*{-0.15cm} \centering
{\hspace*{-0.25cm} \includegraphics[angle=0,width=80mm]{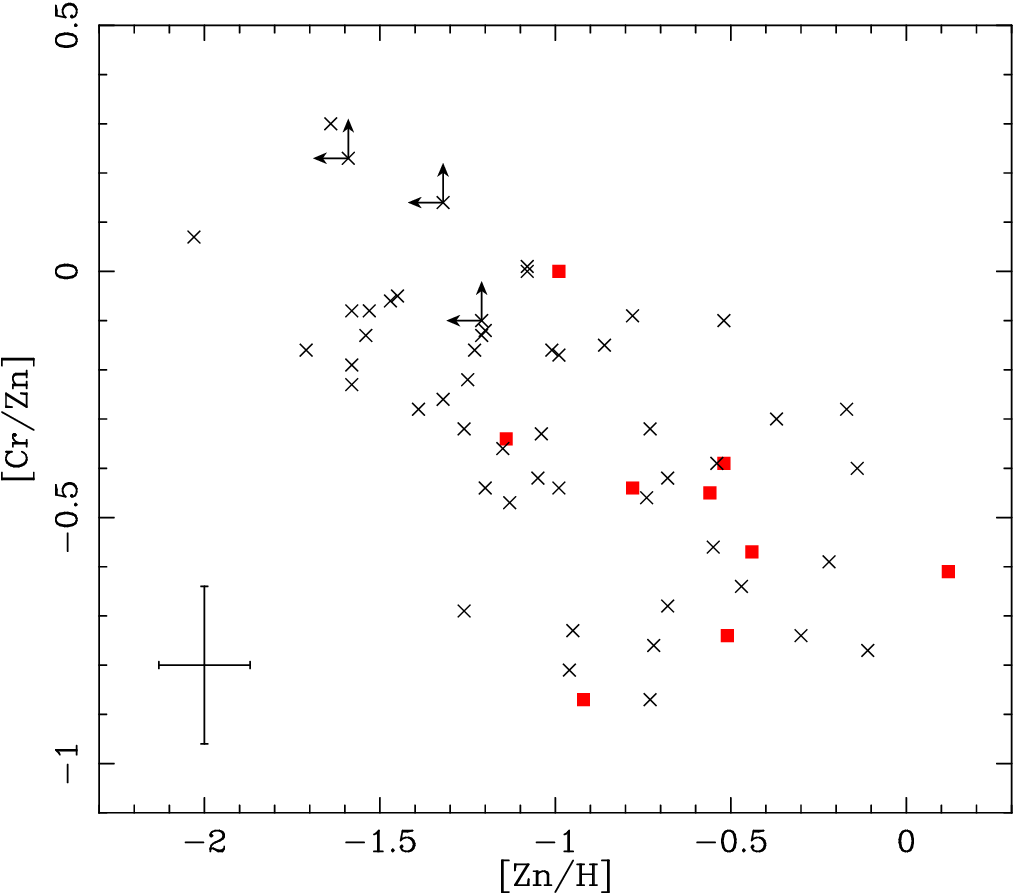}}
  \caption{Measurements of [Cr/Zn] in DLAs  from the compilation by
Kulkarni et al. (2007) and from the present work.  DLAs from the
present work are shown by red squares unless a superior measurement
exists in the literature.  The error bars in the lower left-hand
corner show  the typical uncertainties in the ratios plotted.  }
  \label{Fig:crzn_znh}
\end{figure}

 In Figure~\ref{Fig:crzn_znh} we have plotted 
 [Cr/Zn] as a function of DLA metallicity (as measured by the 
 [Zn/H] ratio), again combining the systems in the compilation
 by Kulkarni et al. (2007) with the new, or improved, measurements
 from the present work.
 Both Zn and Cr track the abundance of Fe in Galactic stars
 with [Fe/H]\,$\simgt -2$ (i.e. in the same metallicity regime
 as the vast majority of DLAs---see Cayrel et al. 2004 and Nissen et al. 2007).
 On the other hand, in the ISM near the Sun, 
 Cr is among the most depleted
 elements with only a small fraction in the gas phase while most of
 it is `hidden away' in dust grains; Zn, on the contrary, exhibits at most
 only mild depletions (Savage \& Sembach 1996). 
 Thus, the [Cr/Zn] ratio in DLAs is a convenient way to assess
 the degree to which refractory elements are depleted in the
 ISM of the galaxies hosting these absorbers (Pettini et al. 1990)
 and, by inference, obtain an indication of their dust content
 (Pettini et al. 1997a).
 
In Figure~\ref{Fig:crzn_znh} we see the clear correlation between
Cr depletion and metallicity that has already been noted 
(e.g. Pettini et al. 1994; Ledoux, Petitjean \& Srianand 2003; 
Akerman et al. 2005; Meiring et al. 2006); 
the data reported here fit the known trend and improve the statistics 
at the high metallicity end. Although the measurements
scatter about the mean relation between Cr depletion and
Zn abundance, the data in Figure~\ref{Fig:crzn_znh} show
that in DLAs with metallicities [Zn/H]\,$\simlt -1.5$, Cr suffers
little depletion relative to Zn; on the other hand, in the DLAs with near-solar
metallicities, $\sim 75$ per cent of the Cr is presumably incorporated
into dust ([Cr/Zn]\,$\simeq -0.6$). This is still a relatively
modest degree of depletion compared to that exhibited
by Cr in diffuse clouds in the Milky Way disk, and more akin
to the values seen along sight-lines which sample mostly
the interstellar medium in the halo of our Galaxy (see Figure 6
of Savage \& Sembach 1996).
The origin and implications of this metallicity-dependent 
depletion pattern have been considered by Vladilo (2002).

\section{Ca\,II Absorption in DLAs}
\label{sec:caii_in_dlas}

A primary aim of the observations presented in this
paper is to establish the Ca\,{\sc ii} content of known
DLAs at intermediate redshifts. Ultimately, we hope to
clarify the relationship between absorbers with
strong  Ca\,{\sc ii} lines, which can be selected from 
ground-based QSO spectra such as those from the
SDSS without reference to the \lya\ line, 
and the more general population of DLAs.  

For the reasons discussed by WHP06, we suspect
that strong \caii\ absorbers may be a subset of the DLA population.  
However, the properties of DLAs that lead 
to strong \caii\ absorption are not clear.  
WHP06 considered three possibilities; they proposed that 
strong \caii\ absorbers select DLAs with:  (i) the
largest \nhi\ values; (ii) the highest metallicities; or (iii) the
largest volume densities,  $n$(H$_{\rm TOT}$).  
A further possibility is an absence or destruction of dust grains, 
leading to unexpectedly large fractions of Ca atoms in the gas 
phase (although this would be in contradiction to the finding
that strong \caii\ absorbers induce
higher-than-average reddening in the background quasar).  
We now consider these possibilities in turn, in light of the 
new data presented here.

\subsection{\caii\ as a Tracer of Column Density?}

\begin{figure}
   \centering
 {\hspace*{-0.0025cm} \includegraphics[angle=0,width=80mm]{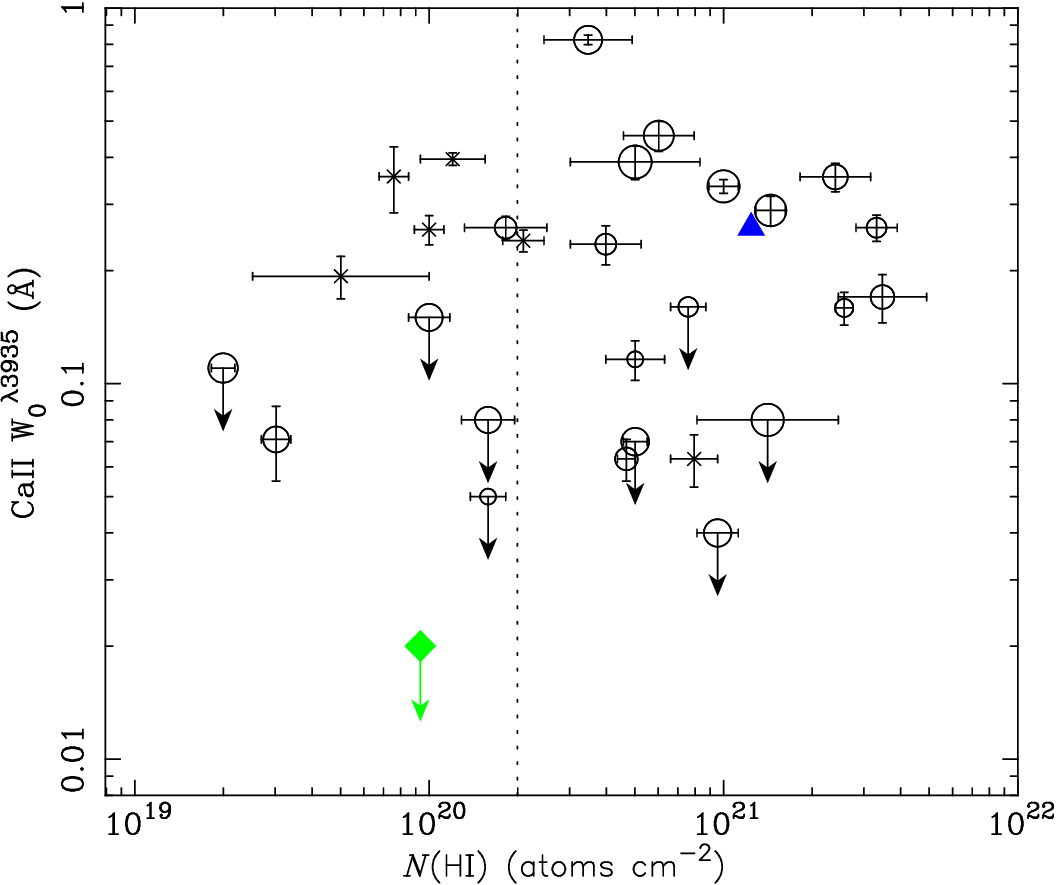}}
  \caption{The rest-frame equivalent width of the Ca\,{\sc ii}\,$\lambda 3935$
  line plotted as a function of the column density of neutral hydrogen.
  \textit{Open circles:} DLAs from the present survey; the size of the
  circle is proportional to the equivalent width of Mg\,{\sc ii}\,$\lambda 2796$
  (0.7\,\AA\,$<$\,\wmg\,$< 3.1$\,\AA---
  see Table~\ref{table:data}). Upper limits to \wca\ are $3 \sigma$. 
  \textit{Crosses:} additional measurements from the literature
  (listed in Table~\ref{table:others}). Note, however, that these are likely
  to be biased towards higher-than-average values of \wca.
  \textit{Filled (blue) triangle and filled (green) diamond:} values measured from
  the stacked spectra of, respectively, the DLAs and sub-DLAs
  in Table~\ref{table:data}, plotted at the corresponding mean values 
  of \nhi\
  (see section~\ref{sec:data_redux}).
  The vertical dash line shows the conventional distinction
  between DLAs and subDLAs at $N$(H\,{\sc i})\,$= 2 \times 10^{20}$\,cm$^{-2}$.
}
  \label{Fig:cah}
\end{figure}

We examine in Figure~\ref{Fig:cah} the relationship between 
the equivalent width of the stronger member of the 
Ca\,{\sc ii} doublet, \wca, and the column density
of neutral hydrogen $N$(H\,{\sc i}). 
We have augmented our sample with a trawl of the literature
for other measurements of Ca\,{\sc ii} in DLAs and subDLAs, collected in
Table~\ref{table:others}. However, it must be borne in mind that
the six additional systems thus found do not represent a statistical 
sample, because non-detections of \caii\ and the corresponding
upper limits on \wca\ generally go unreported in the literature. 

The data collected in Figure ~\ref{Fig:cah} do not support
the hypothesis that strong Ca\,{\sc ii} absorption corresponds exclusively
to DLAs at the
upper end of the $N$(H\,{\sc i}) distribution.
WHP06 showed that, if this were the case, the incidence
of strong \caii\ absorbers per unit redshift would match
that of DLAs with $10^{21} \la$ \nhi $\la 10^{22}$ so that,
in this scenario, the former would account for all of the latter.
In contrast, our combined sample includes 
eight DLAs with \nhi\,$\ge 10^{21}$,
and \emph{none} of them is a strong \caii\ absorber. 

More generally, 
inspection of Figure~\ref{Fig:cah} shows that there is no 
simple relation between \wca\ and \nhi\ in DLAs. 
Selecting absorption systems by Ca\,{\sc ii} equivalent width
does, however, primarily select DLAs: 
among the systems we have observed, all those with 
\wca\,$> 0.15$\,\AA\ are DLAs [within the accuracy
of the determination of $N$(H\,{\sc i})].
On the other hand, there are subDLAs with 
\wca\,$> 0.15$\,\AA\ reported in the literature.
Conversely, selecting by \nhi\ does not ensure that a 
strong \caii\ line is present: 
over one third of the DLAs in the our sample have 
\wca\,$< 0.15$\,\AA\ and three quarters have \wca\,$< 0.3$\,\AA.
On the basis of our composite spectrum 
(blue triangle in Figure\,\ref{Fig:cah} --- see section 2.3),
we conclude that Ca\,{\sc ii} absorption with
an average $\langle {\rm W}_0^{\lambda 3935} \rangle = 0.26\,$\AA\
is associated with known DLAs.
Furthermore, despite the few strong detections 
reported in the literature, the composite spectrum of
the sub-DLAs in our sample shows that 
their associated \caii\ absorption is weak, with
an average $\langle {\rm W}_0^{\lambda 3935} \rangle < 0.02\,$\AA\
(green diamond in Figure~\ref{Fig:cah}).

We note that only four out of the 19 DLAs in the
combined sample meet the selection criterion 
\wca\,$\geq 0.35$\,\AA\
of the SDSS survey by WHP06, and only \emph{one} out of 19
would be classed as a `strong' Ca\,{\sc ii} absorber
(\wca\,$\geq 0.5$\,\AA) by WHP06.
Evidently, there is little overlap as yet between
confirmed DLAs and the SDSS candidates identified
via Ca\,{\sc ii}. 
This is not surprising, however, when we consider
the magnitude limit of the \textit{HST} observations which
RTN06 used to assemble 
the current sample of intermediate redshift DLAs.

Specifically, WHP06 estimate that the true (unbiased) redshift path density of
Ca\,{\sc ii} with \wca\,$>0.5$\,\AA\ is $\sim$ 20--30 per cent of that of DLAs.
However, for a flux-limited sample in the SDSS $i$--band they also calculate
that $\sim 40$ per cent of their strong \caii\ systems are missed due to the
relatively strong extinction of the background QSOs by such absorbers.  A
detailed description of the calculation is provided in Section 6.1 of WHP06.
The majority of quasars included in the RTN06 compilation were discovered either
using much bluer passbands (usually $B$ or $V$), or from an SDSS $g$-band
flux-limited sample.  The fractional loss of strong \caii--systems from the
RTN06 compilation may be calculated using exactly the same procedure as
described in WHP06 but using observed-frame wavelengths appropriate for the
$B$--, $V$-- or $g-$bands instead of the $\sim$ 8000 \AA\ SDSS i-band.  The
significantly shorter absorber rest-frame wavelengths probed by the bluer
passbands mean that the fraction of background QSOs falling below a pre-defined
flux-limit increases to $\sim 80$ per cent, compared to the $\sim 40$ per cent
loss for the SDSS $i$--band.  Assuming that $N$(H\,{\sc i}) and \wca\ are
unrelated in DLAs, we thus calculate that our sample of 16 DLAs was likely to
include only one \caii-absorber with \wca\,$>0.5$\,\AA, as indeed turned out to
be the case (the \wca\,$= 0.82$\,\AA\ absorber at $z_{\rm abs} = 0.7156$ towards
$1323-0021$).  Estimates of the degree of reddening for individual absorbers in
our sample are given in Appendix A.

Returning to Figure~\ref{Fig:cah}, it can also be appreciated that
there is no obvious correlation between \wca\ and \mgii\ \wmg\, --
the size of the open circles in the Figure is proportional to
the latter quantity.  \mgii\ absorption in DLAs is typically
at least moderately saturated and, thus, primarily a tracer
of the velocity structure of the absorbing gas. 

\subsection{\caii\ as a Tracer of Metallicity or Dust Content?}

\begin{figure}
   \centering
 {\hspace*{-0.0025cm} \includegraphics[angle=0,width=80mm]{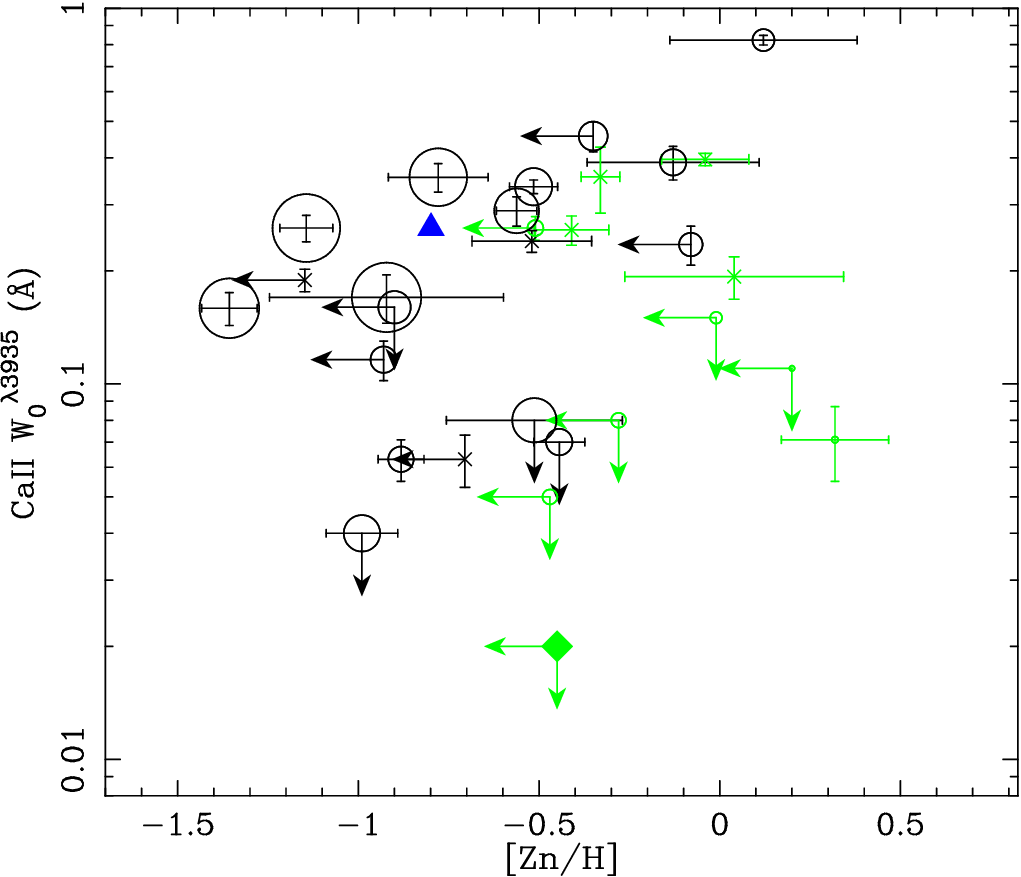}}
  \caption{The rest-frame equivalent width of the Ca\,{\sc ii}\,$\lambda 3935$
  line plotted as a function of DLA metallicity. The symbols have the same
  meaning as in Figure~\ref{Fig:cah}, except that the size of the open circles
  is now proportional to \nhi. SubDLAs are shown in green; for some
  of these, the assumption 
  $N$(Zn\,{\sc ii})/$N$(H\,{\sc i})\,=\,Zn/H adopted in
  section~\ref{sec:abundances} may not
  be correct, possibly leading to an overestimate of the
  metallicity if some of the  Zn\,{\sc ii} is associated with
  ionised gas.
  }
  \label{Fig:caznh}
\end{figure}

We have searched for other clues to
the origin of strong Ca\,{\sc ii} absorbers by testing for
correlations of \wca\ with DLA metallicity (as measured
by [Zn/H]) and dust depletions (via [Cr/Zn], as explained
in Section~\ref{sec:crzn} above). 
As can be seen from Figure~\ref{Fig:caznh},
no such correlation with [Zn/H] is evident in the present data set.
The only DLA in our sample with \wca\,$> 0.5$\,\AA\ is indeed
the most metal-rich known at any redshift\footnote{This system is discussed by 
P{\'e}roux et al. (2006b) as a subDLA based on the $N$(H\,{\sc i}) measurement
of Khare et al. (2004).  However, we use the subsequent measurement by Rao, Turnshek
and Nestor (2006) of $N$(H\,{\sc i})$= 20.54 \pm 0.15$.}, 
with [Zn/H]\,$=+0.12 \pm 0.26$
(see Figures~\ref{Fig:caznh} and \ref{Fig:znhz}).
However, DLAs with \wca\,$ \simgt 0.15$\,\AA\ 
span essentially the full range of values of [Zn/H].
Due to the dearth of strong \caii\ absorbers in our sample,
it is difficult to test the hypothesis that the strongest \caii\
absorbers represent the DLAs with the highest metallicities.
However, the highly-enriched nature of the lone 
strong \caii\ absorber in our survey suggests 
that metallicity may play some role.

\begin{figure}
\centering
  \includegraphics[angle=0,width=80mm]{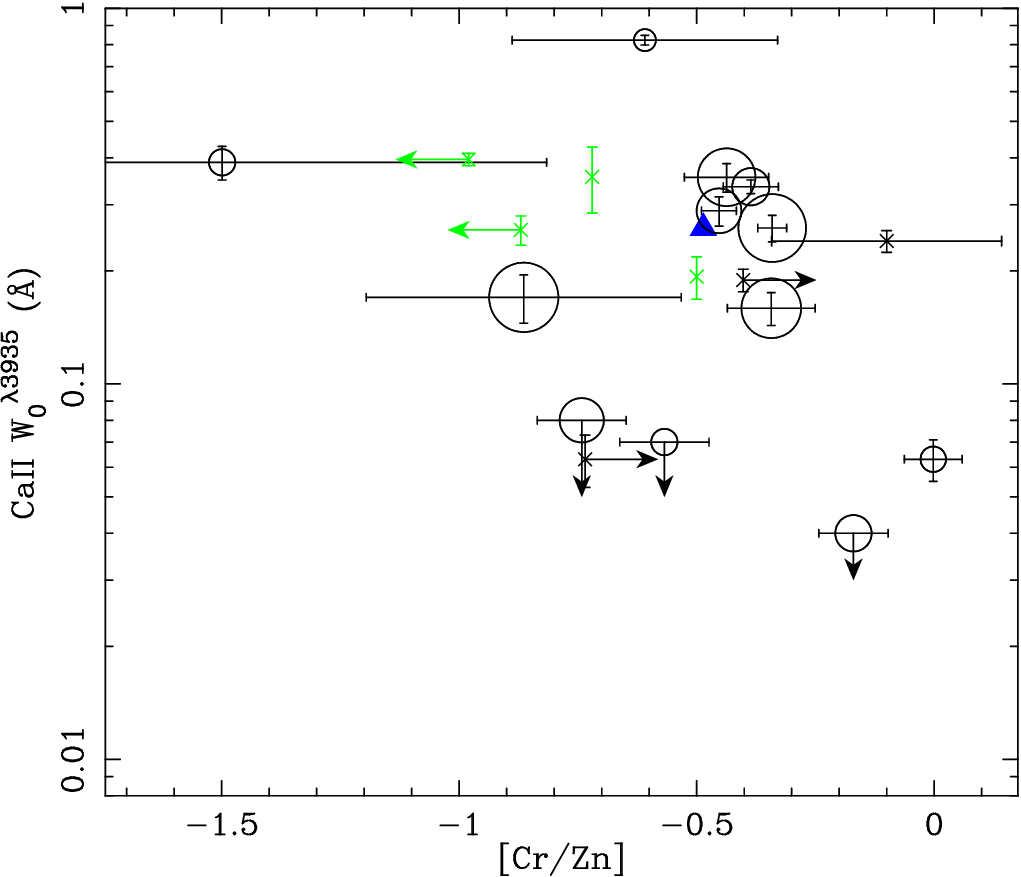}
  \caption{The rest-frame equivalent width of the Ca\,{\sc ii}\,$\lambda 3935$
  line plotted as a function of Cr depletion. The symbols have the same
  meaning as in Figure~\ref{Fig:caznh}.
  }
  \label{Fig:cacrzn}
\end{figure}

Given the highly refractory
nature of Ca, we also consider whether 
a lack of dust grains, leading to a large
fraction of Ca remaining in the gas phase, 
may be an important factor in determining 
\caii\ absorption strength.
However, Figure~\ref{Fig:cacrzn} 
shows no obvious
relationship between \caii\ absorption and depletion of Cr
present.  The values of [Cr/Zn] determined here
are similar to those of other DLA samples and,
in particular, we find no evidence for higher
values of \wca\ in DLAs with low dust depletions
(as evidenced by near-solar values of [Cr/Zn]).
The lone strong \caii\ absorber in our sample
has a [Cr/Zn] value consistent with the average
for systems of similar \caii-strength reported 
by WHP06.

\subsection{\caii\ as a Tracer of Volume Density?}

The strengths of Mg\,{\sc ii}, Zn\,{\sc ii}, and Cr\,{\sc ii}
lines in DLAs are determined primarily by the H\,{\sc i}
column density, metallicity, and dust depletion, as well
as the velocity dispersion of the gas and the oscillator strengths
of the transitions. 
This is because these ions are the dominant species of
Mg, Zn, and Cr in H\,{\sc i} regions, where photons
with sufficient energy to ionise these elements beyond
their first ions do not penetrate.
The same is not true of Ca\,{\sc ii}. With an ionisation
potential of 11.87\,eV, a significant fraction of Ca
can be doubly ionised even when the bulk of the
gas is neutral. In photoionisation equilibrium,
the balance between doubly and singly ionised Ca
is given by:
\begin{equation}
\label{eq:ca}
\frac{n{\rm (Ca^{2+})}}{n{\rm (Ca^{+})}} =
\frac{\Gamma}
{\alpha (T) \, n{\rm (e)}}
\end{equation}
where $n$ denotes volume densities, $\Gamma$ is the 
Ca$^+$ photoionisation rate, $\alpha (T)$ is
the recombination coefficient to form Ca$^{+}$,
and $n{\rm (e)}$ is the electron density.
Presumably, this dependence of $n$(Ca$^+$)/$n$(Ca$_{\rm TOT}$) 
on the local physical conditions is the extra factor responsible for
the lack of any obvious correlation of \wca\ with 
\hi\ column density, metallicity and depletion levels.

We can exploit this feature of Ca\,{\sc ii} absorbers
to gain information on the physical
conditions which determine the ionisation balance
of Ca.  Since we know, or have
estimates of, $N$(H\,{\sc i}),  metallicity (from [Zn/H]),
and dust depletion (from [Cr/Zn]), we can 
use the measured values of $N$(Ca\,{\sc ii}) 
-- determined via \wca\ --
to estimate the ratio
$n{\rm (Ca^{2+})}/n{\rm (Ca^{+})}$.  
For a range of values of the intensity of the far-UV radiation
field (which determines the photoionisation rate $\Gamma$)
we can then deduce the corresponding values of  
gas density $n$(H$_{\rm TOT}$) (on which $n$(e)
depends), and thereby assess the hypothesis
the gas volume density is the important factor 
regulating the strength of \caii\ absorption.  
This, in turn, may shed light on the nature
of the strong \caii\ absorbers and their relationship, for
example, to DLAs with associated molecular hydrogen
absorption (Srianand et al. 2005).

\subsubsection{Photoionisation models}

With this aim, we ran series of photoionisation models with the 
software package {\sc cloudy}\footnote{http://www.nublado.org/}
(Ferland et al. 1998; Ferland 2000),
approximating the DLAs as slabs of constant density gas
irradiated by the metagalactic ionising background (Haardt \& Madau 2001)
and the cosmic microwave background at the 
appropriate redshift. We further added cosmic rays and
an interstellar radiation field due to the OB
stars within the galaxies giving rise to the DLAs; 
the spectral distribution of this last component
was assumed to be the same as that of the 
interstellar radiation field in the solar
vicinity, as given by Black (1987).
For each DLA in Tables~\ref{table:data} and \ref{table:others}
where Zn\,{\sc ii} and Cr\,{\sc ii} lines were detected,
we adopted the metallicity implied by the measures of [Zn/H]
(thereby assuming the dust depletion of Zn to be negligible in DLAs) and
included dust grains with an abundance scaled with
the DLA metallicity.  We explicitly accounted for the depletion of metals onto 
grains by assuming solar relative abundances as given by Lodders (2003),
and a depletion pattern motivated by the well-studied interstellar cloud in front
of the star $\zeta$~Oph (Savage \& Sembach 1996).  We scaled the level of 
depletion for each element more highly refactory than Zn based on our
[Cr/Zn] measurement:

\begin{equation}
\label{eq:cacr}
{\rm [X/Zn]}_{\rm gas} = A({\rm X}) \times {\rm [Cr/Zn]}_{\rm gas},
\end{equation}
where $A({\rm X})$ is determined from the depletion values given by 
Savage \& Sembach.  
This functional form matches the relative degrees of
depletion from the gas-phase in the ISM 
towards $\zeta$~Oph, while
reducing to [X/Zn]$_{\rm gas} = 0$ when [Cr/Zn]$_{\rm gas} = 0 $
(i.e. when Cr is undepleted, so are all other metals).  

Each model was characterised by a pair of values: gas
density and intensity of the external stellar radiation field.  
We then used {\sc cloudy} to predict the value of $N$(Ca\,{\sc ii})
appropriate to that model given the measured $N$(H\,{\sc i}) and all
of the above parameters. We assigned a $\chi^2$ confidence level to
each model, by comparing predicted and observed values of $N$(Ca\,{\sc
ii}) and taking into account the uncertainties in our measurements of column
densities.   We assumed an uncertainty of $\pm 0.3$ applied to the
scaling factor in eq.(\ref{eq:cacr}) for Ca; i.e., $A({\rm Ca}) = 1.9 \pm 0.3$.  
Although {\it ad hoc}, it is likely a 
conservative estimate.  The relative contribution to the total
error arising from this uncertainty depends on the level  of
depletion, and ranged between zero and 50 per cent, with an average of
25 per cent, in our models.  Thus, for each DLA in Tables~1 and 2
with measured \nzn\ and \ncr\, we were able to produce confidence
intervals in gas density (cm$^{-3}$) for a given choice of intensity of
the interstellar radiation field (in units of the local ISM radiation
field intensity).   

\begin{figure}
\centering
{\hspace*{-0.25cm} \includegraphics[angle=0,width=80mm]{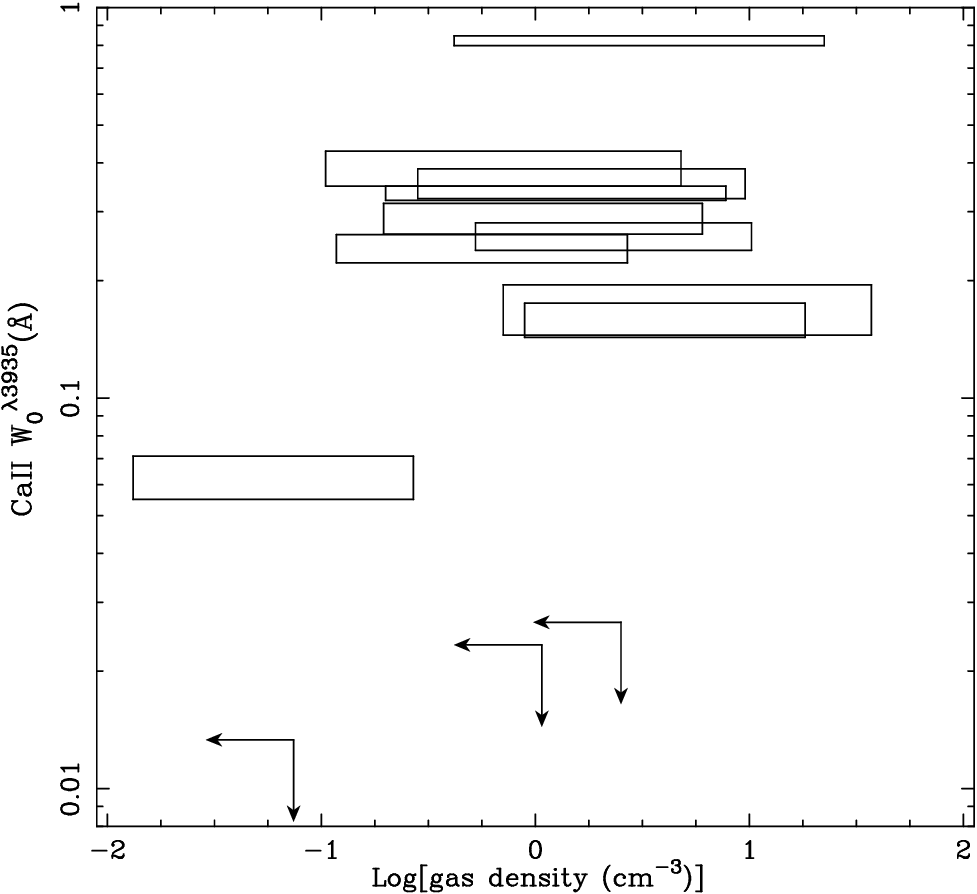}}
  \caption{The rest-frame equivalent width of the Ca\,{\sc ii}\,$\lambda 3935$
  line plotted as a function of the values of $n$(H$_{\rm TOT}$)
  deduced from our {\sc cloudy} modelling. The range of values
  of $n$(H$_{\rm TOT}$) shown corresponds to the full range
  of normalisations of the UV radiation field considered in our
  modelling, from 1/10 to three times the local interstellar value.
  Upper-limits for \wca\ and $n$(H$_{\rm TOT}$) are shown at one-sigma.
 }
  \label{Fig:hden_range}
\end{figure}

Figure~\ref{Fig:hden_range} shows the $1 \sigma$ ranges 
in gas density if the intensity of the interstellar
radiation field is allowed to vary between 1/10 and three
times the reference value appropriate to our location
within the Milky Way. 
We see that, even in the best cases, we only constrain 
$n$(H$_{\rm TOT}$) to within $\approx$ 1--2 
orders of magnitude, depending
on the level of star formation activity in the 
galaxies hosting the DLAs (since the star formation rate
is presumably the main determinant of the radiation field intensity). 
Previous estimates of the interstellar radiation field intensity
in DLAs have been deduced from consideration of the
relative populations of the rotationally excited levels
of the H$_2$ molecule (e.g. Ge \& Bechtold 1997).
Available data indicate a UV field intensity
comparable to the local value (Srianand et al. 2005),
consistent with the conclusion from most 
imaging studies that the galaxies giving
rise to DLAs are generally fainter than $L^{\ast}$
(Rao et al. 2003 and reference therein).

\begin{figure}
\centering
{\hspace*{-0.25cm} \includegraphics[angle=0,width=80mm]{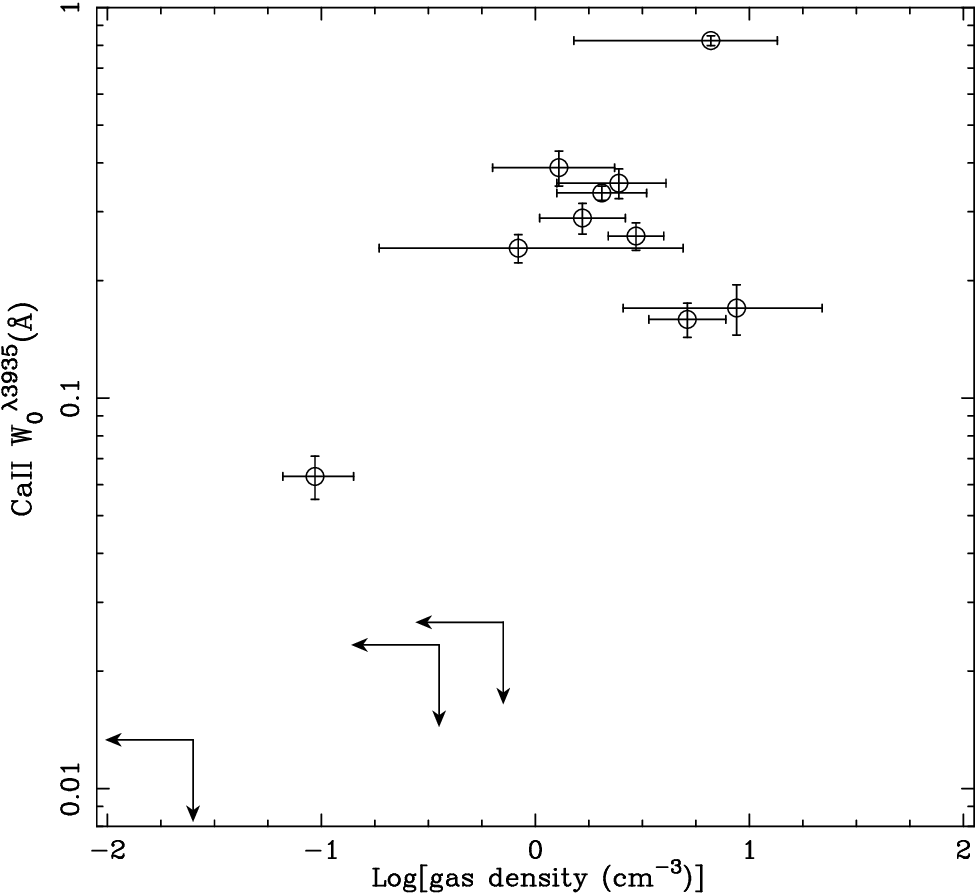}}
  \caption{As for Figure~\ref{Fig:hden_range}, but now assuming
 that the interstellar radiation fields in the galaxies hosting the DLAs 
 have the same intensity as that of the Milky Way near the Sun.
 }
  \label{Fig:hden_MW}
\end{figure}

These measurements refer to DLAs
at redshifts $z \simgt 2$ and it is thus unclear to
what extent they also apply to lower redshift
systems. If we assume, as a working assumption,
a UV field equal to that observed locally, we obtain
the values of gas density shown in Figure~\ref{Fig:hden_MW}
and listed in Table~\ref{table:nTOT} 
(where the DLAs are now listed in increasing
order of \wca).  The range of \wca\ in our sample is too small, 
and in some cases 
$n$(H$_{\rm TOT}$) is too poorly constrained,
to discern an overall correlation between
\wca\ and $n$(H$_{\rm TOT}$). 
However, we note that in the DLA with the lowest measured
value of \wca\ (at $z_{\rm abs} = 0.8591$
in 0454$+$039, with \wca\,$= 0.063$\,\AA)
we deduce the lowest value of gas density,
$n$(H$_{\rm TOT}) \simeq 0.1$\,cm$^{-3}$.
Such a weak Ca\,{\sc ii} line cannot be ascribed
to dust depletion of Ca, since in this DLA
the Cr/Zn ratio is solar.
Conversely, the only `strong' \caii\ absorber
in our sample, according to the definition by WHP06
(at $z_{\rm abs} = 0.7156$
in 1323$-$0021, with \wca\,$= 0.82$\,\AA),
requires relatively high density according to our modelling,
with $n$(H$_{\rm TOT}) \simeq 7$\,cm$^{-3}$.
In this, as in other DLAs with large values of \wca,
we may have underestimated $N$(Ca\,{\sc ii}), and therefore
$n$(H$_{\rm TOT})$ (by $\sim$ 0.1 - 0.5 dex), 
if there is hidden saturation in the absorption lines.

We can apply a similar analysis to the `low-\wca' 
($\langle {\rm W}_0^{\lambda 3935}\rangle = 0.49$\,\AA)
and `high-\wca' ($\langle {\rm W}_0^{\lambda 3935}\rangle = 1.01$\,\AA)
subsamples of \caii\ absorbers from WHP06, for which 
those authors measured values of \ncaii, \nzn, and \ncr\
from composites of SDSS spectra.  Since we do not know
the value of \nhi\ appropriate to these composites,
we have performed the calculation for a range of
values of the neutral hydrogen
column density (and therefore also a range
of metallicities).  For typical DLA \nhi-values, we find that, 
for a normalisation of the UV field equal to the local value, 
the typical strong \caii\ absorber arises in gas with density
in the range $n$(H$_{\rm TOT}) \simeq 10^{0.8} - 10^{1.3}$\,atoms~cm$^{-3}$ 
for the low-\wca\ subsample, and
$n$(H$_{\rm TOT}) \simeq 10^{1.1} - 10^{1.6}$\,atoms~cm$^{-3}$, 
for the high-\wca\ subsample,
depending on the value of \nhi.

\subsubsection{Connection with H$_2$?}

The current evidence thus suggests that strong \caii\ absorbers
trace gas of relatively high density.
Returning to our sample, it can be seen from 
Figure~\ref{Fig:hden_MW} and Table~\ref{table:nTOT}
that if the UV radiation field intensity is comparable to
that in the solar neighbourhood, most Ca\,{\sc ii}-bearing DLAs, 
and all four with \wca\,$ \simgt  0.35$\,\AA, 
likely arise in gas with densities greater than  $\sim 1$ atom~cm$^{-3}$,
while for the average \caii\ system of WHP06 we find 
densities $\ga 10$ atom~cm$^{-3}$.
These values approach the range derived by Srianand et al. (2005) 
in DLAs at $z > 2$ also exhibiting H$_2$ absorption,
from consideration of the relative populations
of the fine-structure levels of the ground state of C\,{\sc i}.
These authors propose that such systems
trace the `Cold Neutral' phase of the interstellar
medium in galaxies; our study raises the possibility
that absorption systems selected from their
strong Ca\,{\sc ii} absorption may be the
counterpart of such DLAs at lower redshift.
In the last column of Table 
\ref{table:nTOT}, we list the molecular fraction, 
$f_{\rm H_2} = 2$N(H$_2$)/(2\,N(H$_2) + $\nhi), 
predicted by our CLOUDY models for a normalization 
of the local UV field of unity.  The range of 
values is broadly consistent with those found by
Noterdaeme et al.\ (2008) for DLAs at redshifts 
$2 \la z \la 4$, although we predict a smaller fraction
of very low $f_{\rm H_2}$-values.  
However, the lowest $f_{\rm H_2}$-values in the Noterdaeme et al.\ sample
are dominated by the systems with the lowest metallicity and 
depletion, both of
which are (on average) greater at the redshifts of our sample
(Figures \ref{Fig:znhz} and \ref{Fig:crzn_znh}).  If we consider
only systems with metallicities greater than
$\simeq -1.3$ in the high-$z$ Noterdaeme et al.\ sample, the values
predicted by CLOUDY are in reasonable agreement with
observed molecular fractions.  For DLAs with properties similar
to the low--\wca\ and high-\wca\ subsamples of WHP06, our models predict 
$f_{\rm H_2} \simeq -2.8$ to $-2.1$ and $f_{\rm H_2} \simeq -0.4$, respectively.
If these interpretations are correct, 
we expect that future space-based observations
of strong Ca\,{\sc ii} absorbers at far-UV wavelengths
will show not only a damped \lya\ line,
but also detectable levels of H$_2$ absorption.

\begin{table}
\centering
\begin{minipage}[c]{1.0\textwidth}
\caption{\textsc{Predicted gas densities and molecular fractions$^{\rm a}$
}}
\begin{tabular}{@{}lccc}
\hline
   \multicolumn{1}{c}{~~~QSO~~~~~~}
& \multicolumn{1}{c}{~~~~~~\wca~~~~~~}
& \multicolumn{1}{c}{~~$\log n$(H$_{\rm TOT}$)~~}
& \multicolumn{1}{c}{$\log f_{\rm H_2}$}\\
   \multicolumn{1}{c}{}
& \multicolumn{1}{c}{(\AA)}
& \multicolumn{1}{c}{(cm$^{-3}$)}
& \multicolumn{1}{c}{} \\
\hline
~~~0449$-$168$^{\rm b}$  &$< 0.03 $                   & $ < -1.6$   & $< -9.4 $    \\
~~~1733+5533$^{\rm b}$   &    $< 0.02 $               & $ < -0.45$   & $< -6.6 $    \\
~~~1007+0042$^{\rm b}$   & $<0.03$                    & $ < -0.14$   & $< -5.5 $     \\
~~~0454+039     & $0.06 \pm 0.01$           & $  -1.03^{+0.18}_{-0.15} $  & $-8.5$  \\
~~~1727+5302   & $0.16 \pm 0.02$            & $  0.71 \pm 0.18$   & $-3.9$  \\
~~~2353$-$0028    & $0.17 \pm 0.03$         & $  0.94^{+0.41}_{-0.53} $  & $-1.2$ \\
~~~2328+0022    & $0.24 \pm 0.02$           & $ -0.08^{+0.77}_{-0.65} $  & $-6.3$   \\
~~~1010+0003   & $0.26 \pm 0.02$            & $  0.47 \pm 0.13$ & $-3.6$ \\
~~~1727+5302   & $0.29 \pm 0.03$            & $  0.22 \pm 0.20$ & $-3.1$ \\
~~~1107+0048   & $0.34 \pm 0.01$            & $  0.31 \pm 0.21$ & $-3.1$ \\
~~~1225+0035   & $0.36 \pm 0.03$            & $  0.39^{+0.22}_{-0.29} $ & $-2.6$ \\
~~~0256+0110$^{\rm c}$   & $0.39 \pm 0.04$   & $  0.11^{+0.26}_{-0.31}  $ & $-3.3$ \\
~~~1323$-$0021    & $0.82 \pm 0.02$         & $  0.82^{+0.31}_{-0.64} $  & $-0.6$    \\
\hline
\end{tabular}
\\
$^{\rm a}${Assuming\, that the intensity of the stellar UV radiation\, field in \\
the DLAs is similar to the mean value measured in the\\
interstellar medium of the Milky Way near the Sun.  Upper-limits\\ 
correspond to one-sigma confidence bounds.}\\
$^{\rm b}$ One-sigma upper-limits on \wca\ were used in the models.\\
$^{\rm c}${We use the $\log N$(Cr\,{\sc ii}) value
given in the literature for\\ 
modelling the DLA towards 0256+0110 as our measurement\\
is highly uncertain.}\\
\label{table:nTOT}
\end{minipage}
\end{table}

At present, we can test this prediction with only 
three reported measurements of,  or limits on, $N$(H$_2$)
at $z_{\rm abs} < 1.3$: in the DLAs towards
0454+039 (which is in our sample) and 1328+307 
($z_{\rm abs}=0.692$; $N$(\hi)\,$=10^{21.28}$\,cm$^{-2}$), 
and in the subDLA towards 0515-4414 
($z_{\rm abs} = 1.1508$;
$N$(H\,{\sc i})\,$= 10^{19.88}$\,cm$^{-2}$ -- see Table 2.)  
Molecular hydrogen is undetected
in either of the two DLAs (Ge \& Bechtold 1999), with 
$N$(H$_2) < 10^{16.38}$ and $f_{\rm H_2} < 10^{-4.01}$ for the 
system towards $0454+039$ and  $N$(H$_2) < 10^{15.70}$ and 
$f_{\rm H_2} < 10^{-5.28}$ for the system towards $1328+307$.
The former has the weakest detected \caii\ line in our sample.
For the latter, we estimate \wca $\simlt 0.2$\,\AA\
from its SDSS spectrum.
Thus, neither of these two DLAs are strong Ca\,{\sc ii}
absorbers. 
In contrast, the subDLA towards 0515-4414
does exhibit H$_2$ absorption 
(Reimers et al. 2003),
with $N$(H$_2) = 10^{16.94 \pm 0.30}$ and 
$f_{\rm H_2} = 10^{-2.64 \pm 0.30}$.
In this system, \wca $=0.36$\,\AA;
while this does not qualify it as a `strong' \caii\ absorber
in the sense defined by WHP06, we note that this value
of \wca\ is unusually high considering the low \nhi\
(see Figure~\ref{Fig:cah}).
For comparison, in the lone strong \caii\ system in our sample,
\wca\ is greater only by a factor of about two,
despite having column densities
of Zn\,{\sc ii} and Cr\,{\sc ii} one order of magnitude higher.
It will be of interest to verify the extent to which 
\wca\ and $N$(H$_2$) are related as more measurements
of these quantities become available.

\section{Summary and Conclusions}
\label{Sec:Sum} 
We have used the WHT telescope on La Palma
to record at intermediate resolution the optical spectra
of QSOs known to lie behind DLAs and subDLAs in
the redshift range $0.6 < z_{\rm abs} < 1.3$. 
Our principal aims were twofold: (i) to measure the strength 
of the Ca\,{\sc ii}\,$\lambda \lambda 3935, 3970$
absorption lines in known DLAs, 
with a view to establishing to what extent
DLAs and strong Ca\,{\sc ii} absorbers overlap; and 
(ii) to measure the 
\znii\,$\lambda\lambda 2026, 2063$
and \crii\,$\lambda\lambda 2056, 2062, 2066$ 
multiplets in the (sub)DLAs, thereby increasing the 
number of metallicity and depletion determinations
in (sub)DLAs at low-redshift.
By design, these complementary measurements 
allow us to test whether the strength of Ca\,{\sc ii} absorption
is related to the metallicity of the gas, as measured
by the abundance of Zn, and/or to the degree
of dust depletion of refractory elements
(both Ca and Cr are readily incorporated onto
dust, whereas Zn is not).
From consideration of such measurements
in a sample of 16 DLAs and six subDLAs,
augmented by similar data for three DLAs and four subDLAs
from published studies, we reach the
following conclusions.

\begin{itemize}

\item[1.] The new measurements presented here strengthen 
previous conclusions on the metallicities and dust depletions in
DLAs at $z < 1.3$. We point
out that, after accounting for the fact that at these redshifts DLAs
have mostly been selected from Mg\,{\sc ii}-strong systems -- and may
thus be biased in favour of metal-rich absorbers -- the already weak
redshift evolution of the mean  DLA metallicity is reduced further:
[$\langle {\rm Zn/H}\rangle$]
is only a factor of about two greater at  $0.6 < z < 1.3$ 
than at higher redshifts. 
We also confirm the dependence
of dust depletion on metallicity uncovered by earlier studies.

\item[2.] Most DLAs exhibit detectable Ca\,{\sc ii}\,$\lambda 3935$
absorption in our spectra.  Our new data indicate that
selecting absorption systems above the threshold  \wca\,$ =
0.15$\,\AA\ isolates systems with \nhi\,$\ga 2 \times
10^{20}$\,cm$^{-2}$ (although with some contamination from subDLAs,
as evidenced by detections reported in the literature).
However, while it seems likely that 
strong Ca\,{\sc ii} absorption is associated with DLAs,
the converse is not true: approximately one third of the DLAs in our
sample have \wca\,$< 0.15$\,\AA, while adopting the criterion
\wca\,$> 0.30$\,\AA\ would miss $\simeq 3/4$ of the DLAs.

\item[3.] Only four out of the 19 DLAs in the
combined sample meet the selection criterion \wca\,$\geq 0.35$\,\AA\
of the SDSS survey by WHP6, and only one out of 19
would be classed as a `strong' Ca\,{\sc ii} absorber
by their definition (\wca\,$\geq 0.5$\,\AA).
Evidently, there is little overlap as yet between
confirmed DLAs and the SDSS candidates identified
via Ca\,{\sc ii}. This is not surprising, however, 
since -- as demonstrated by WHP06 -- dust associated with 
many of the strong Ca\,{\sc ii} absorbers dims the light from
the background QSOs sufficiently for them to be 
excluded from the magnitude-limited surveys 
conducted with \textit{HST} up to now.

\item[4.] Overall, we find no 
convincing correlation of the strength  of Ca\,{\sc ii}
absorption with the column density of neutral gas,
the equivalent width of Mg\,{\sc ii}\,$\lambda 2796$, 
the metallicity of the gas
(as measured from the abundance of Zn), or
the degree of depletion of refractory elements
(indicated by the [Cr/Zn] ratio).
Despite the fact that, in the interstellar medium of the 
Milky Way, both
Ca and Cr exhibit large and variable depletion
factors from the gas-phase, DLAs with near-solar
values of [Cr/Zn] do not show significantly stronger
Ca\,{\sc ii} lines than those where most of the
Cr is presumably in solid form.
On the other hand, 
the finding that the only `strong' \caii\ absorber
in our sample is the most metal-rich DLA yet
discovered suggests that there may be some
connection between the metallicity and
\caii\ equivalent width.

\item[5.] Presumably, a more complex mix of factors  than simply
H\,{\sc i} column density, metallicity, and dust depletion, determines
the value of \wca\ in DLAs.  We have presented preliminary evidence
that the `joker in the pack' may be the variable fraction of Ca which
is singly ionised in H\,{\sc i} regions, where Ca\,{\sc iii} is
often  the dominant ion stage, depending on the densities of
particles, far-UV photons, and on the temperature of the gas.

With the data at hand, we exploit the ionisation balance of Ca to
place constraints on the ratio  of UV photons and particles.  If we
further assume that the interstellar radiation field in the DLAs is of
the same order as that at our location within the Milky Way -- as indeed
found for DLAs at higher redshifts -- we can arrive at estimates of, or
limits on,  the gas density $n$(H$_{\rm TOT})$.  Our modelling, using
the photoionisation code {\sc cloudy}, indicates values  $n$(H$_{\rm
TOT}) \simgt 1$\,cm$^{-3}$ for most DLAs with \wca\,$ \simgt
0.35$\,\AA.  Furthermore, for an `average' \caii-absorber DLA 
(as described by WHP06) as well as the lone `strong' (as defined by
those authors) \caii\ system in our sample, we deduce
$n$(H$_{\rm TOT}) \simgt 10$\,cm$^{-3}$.  Such high values are
similar to those which seem to be typical of H$_2$-bearing DLAs at $z
> 2$, raising the possibility that in the strong Ca\,{\sc ii}
absorbers  selected by WHP06 at $z < 1.3$ we may be
seeing the lower redshift counterparts of the subset of DLAs with a
measurable molecular fraction.  Indeed, our models predict molecular
fractions similar to those reported for relatively metal-rich DLAs
at high-$z$, with the strongest \caii\ absorbers exhibiting 
values consistent with the highest reported measurements of 
$f_{\rm H_2}$ in DLAs.

\end{itemize}

If borne out by additional studies,
the possible connection of strong \caii\ absorbers to 
metal- and relatively molecule-rich gas has
intriguing consequences when considered in light of the imaging
results of Hewett \& Wild (2007).  It would mean that
strong \caii\ absorbers select gas with these properties not only in 
galactic disks, but also at distances of tens of kpc from relatively bright
galaxies.  Such a conclusion would be highly relevant to models
of galactic outflows and star-formation feedback.

The currently observed population of DLAs are {\it not}, in general,
strong \caii\ absorbers.  However, if  most or all strong
\caii\ absorbers are DLAs and the true incidence of strong \caii\
absorbers is as high as calculated by WHP06, these
intriguing systems may represent a (potentially metal-rich) subset of
DLA absorbers that are being missed from the current surveys.   It
will be possible to make progress towards clarifying this and
other questions raised by the present work with COS measurements of 
metal lines and 
\nhi\ in a sample of {\it known} strong \caii\ absorbers.  Like the rest of
the UV-spectroscopy community, we look forward with anticipation to
the forthcoming \textit{HST} servicing mission that will install this
much needed instrument.

\section*{Acknowledgments} 

The authors wish to thank Bob Carswell ({\sc vpfit}), and Gary
Ferland and Ryan Porter ({\sc cloudy}) for providing, maintaining, and
offering help with their software.  We are grateful to the WHT time
assignment committee for their support of this programme and to the
staff astronomers on La Palma for their help with the observations.
DBN, MP and PCH acknowledge support from the STFC-funded Galaxy Formation and 
Evolution programme at the Institute of Astronomy. 
VW is supported by the MAGPOP Marie Curie EU Research and Training
Network.

\newpage
\begin{appendix}
\section{QSO coordinates, magnitudes, and reddening estimates}
Here we list the celestial coordinates, magnitudes, and estimates of $E(B-V)$ 
for the quasars listed in Tables 1 and 2.  Magnitudes are given for the SDSS 
$i^\prime$ filter, when possible.  $E(B-V)$ for systems in our sample with an available 
SDSS spectrum is estimated exactly as in Wild \& Hewett (2005).  Note that the scatter 
in individual reddening values (due to the intrinsic variation in the quasar SEDs 
from object to object) cause some `negative reddenings' to occur.\\

  \centering
    \begin{tabular}{@{}ccccc}
      \hline
      Name & Ra (J2000) & Dec (J2000) & m$_{i(AB)}$ & $E(B-V)$\\
      \hline 
      0139$-$0023 & 01 39 38.71  & $-$00 23 48.1 & 18.57 & $-0.099$ \\
      0253+0107 & 02 53 16.46 & +01 07 59.8 & 18.56      & $+0.001$ \\
      0256+0110 & 02 56 07.24 & +01 10 38.7 & 18.42      & $-0.031$ \\
      0449$-$168 & 04 52 14.23 & $-$16 40 16.8 & $m_{\mathrm V}=17.0$ & \ldots\\
      0454+039 & 04 56 47.2 & +04 00 53 & $m_{\mathrm V}=16.5$ & \ldots\\
      1007+0042 & 10 07 15.54 & +00 42 58.4 & 18.75      & $+0.029$\\
      1009-0026 & 10 09 30.47 & $-$00 26 19.2 & 17.28      & $-0.118$\\
      1010+0003 & 10 10 18.20 & +00 03 51.2 & 18.03      & $-0.039$\\
      1107+0048 & 11 07 29.04 & +00 48 11.2 & 17.19      & $-0.036$ \\
      1107+0003 & 11 07 36.68 & +00 03 29.6 & 18.04      & $+0.089$\\
      1220$-$0040 & 12 20 37.01 &  $-$00 40 32.4 & 18.33      & $-0.022$\\
      1224+0037 & 12 24 14.30 &  +00 37 09.0 & 18.53      & $-0.052$\\
      1225+0035 & 12 25 56.62 &  +00 35 35.0 & 18.52      & $+0.072$\\
      1323$-$0021 & 13 23 23.79 & $-$00 21 55.3 & 17.64      & $+0.220$\\
      1727+5302 & 17 27 39.03 &  +53 02 29.2 & 18.04      & $-0.028$\\
      1733+5533 & 17 33 22.99 &  +55 33 00.9 & 17.86      & $-0.018$\\
      2149+212 & 21 51 45.9 & +21 30 14 & $m_{\mathrm V}=19.0$ & \ldots\\
      2353$-$0028 & 23 53 21.63 & $-$00 28 40.7 & 18.23      & $+0.025$\\
      & & & \\
      0515$-$4414 & 05 17 07.61 & $-$44 10 56.2 & $m_{\mathrm B}=15.0$      & \ldots\\
      0738+313 & 07 41 10.70 & +31 12 00.2 & 16.66      & \ldots\\
      1436$-$0051 & 14 36 45.06 & $-$00 51 50.5 & 18.25      & \ldots\\
      2328+0022 & 23 28 20.37 & +00 22 38.2 &  17.78      & \ldots\\
      2331+0038 & 23 31 21.82 & +00 38 07.4 &  17.51      & \ldots\\
      2335+1501 & 23 35 44.19 & +15 01 18.4 & 18.22       & \ldots\\
      \hline
    \end{tabular}

\end{appendix}

\end{document}